\documentclass[pra,aps,twocolumn,showpacs,preprintnumbers,amsmath,amssymb,tightenlines,floatfix,superscriptaddress]{revtex4-1}

\usepackage{graphicx}
\usepackage{graphicx}% Include figure files
\usepackage{dcolumn}% Align table columns on decimal point
\usepackage{bm}% bold math
\usepackage{epsfig}
\usepackage{stmaryrd}
\usepackage{enumerate}
\usepackage{color}
\usepackage[normalem]{ulem}

 \newcommand{\kbar}{\hbar_{\textrm{eff}}} 
 \newcommand{\real}[1]{\mathfrak{Re}\{#1\}}
 \newcommand{\imag}[1]{\mathfrak{Im}\{#1\}}
\newcommand{\qe}{\lambda}
\newcommand{\qenl}{E}

\newcommand{\sub}[2]{{#1}_{\mbox{\!\! \scriptsize #2}}}

\def\beq{\begin{align}}
\def\eeq{\end{align}}
\def\nnl{\\[0.15cm] \nonumber}

\def\CR{\nonumber\\[0.15cm]}
\newcommand{\rref}[1]{Ref.~\cite{#1}}
\newcommand{\fref}[1]{Fig.~\ref{#1}}
\newcommand{\frefp}[2]{Fig.~\ref{#1}(#2)}
\newcommand{\bref}[1]{(\ref{#1})}
\newcommand{\eref}[1]{Eq.~(\ref{#1})}
\newcommand{\sref}[1]{Section \ref{#1}}
\newcommand{\tref}[1]{Table \ref{#1}}
\newcommand{\aref}[1]{Appendix \ref{#1}}

\begin{document}

\title{Macroscopic self-trapping in the dynamical tunneling of a Bose-Einstein condensate}
\author{Sebastian W\"uster}
\affiliation{School of Mathematics and Physics, Brisbane, QLD 4072, Australia}
\affiliation{Max Planck Institute for the Physics of Complex Systems, N\"{o}thnitzer Strasse 38, 01187 Dresden, Germany}
\affiliation{Department of Physics, Bilkent University, Ankara 06800, Turkey}
\affiliation{Department of Physics, Indian Institute of Science Education and Research, Bhopal, Madhya Pradesh 462 066, India}
\email{sebastian@iiserb.ac.in}
\author{Joy Cree}
\affiliation{ARC Centre of Excellence in Future Low-Energy Electronics Technologies, University of Queensland, School of Mathematics and Physics, Brisbane, QLD 4072, Australia}
\affiliation{Stanford Institute for Theoretical Physics, Stanford, CA, United States of America}
\author{Matthew J. Davis}
\affiliation{ARC Centre of Excellence in Future Low-Energy Electronics Technologies, University of Queensland, School of Mathematics and Physics, Brisbane, QLD 4072, Australia}

\begin{abstract}
A Bose-Einstein condensate in a modulated, one-dimensional, anharmonic potential can exhibit dynamical tunneling between islands of regular motion in phase space. 
With increasingly repulsive atomic interactions, dynamical tunneling is predicted to cease due to self-trapping [S.~W\"uster {\it et al.} Phys.\ Rev.\ Lett.~{\bf 109} 080401 (2012)]. This suppression of tunneling oscillations 
is related to the same phenomenon that occurs in the two-mode dynamics of a repulsively interacting Bose-Einstein condensate in a double-well potential.  Here we present a two-mode model for dynamical tunnelling based on nonlinear Floquet states and examine the range of validity of the approximation.
We characterise nonlinear dynamical tunneling for different trap strengths, modulation amplitudes, and effective Planck constants. Using the linear Floquet states we derive an expression for the  
critical nonlinearity beyond which tunneling ceases. Finally we demonstrate the dynamical instability of selected nonlinear Floquet states and show how to initialise some Floquet states in experiments. Our detailed survey will enable experiments to target accessible parameter regimes for the study of nonlinear dynamical tunneling.
\end{abstract}

\maketitle

%%%%%%%%%%%%%%%%%%%%%%%%%%%%%%%%
%%%%%%%%%%%%%%%%%%%%%%%%%%%%%%%%
%%%%%%%%%%%%%%%%%%%%%%%%%%%%%%%%
\section{Introduction}
\label{intro}
Particle transport that is classically forbidden may nonetheless take place via quantum-mechanical tunneling.
For example, a particle in a double-well potential cannot classically move from one well to the other if the barrier has a higher potential energy than the total energy of the particle. Dynamical tunneling is a subtly different process that can occur in systems that are periodically modulated in time~\cite{book:reichl}.  Here the classical transport process is inhibited due to the formation of Kolmogorov-Arnold-Moser (KAM) tori of regular motion in phase space from which classical trajectories cannot escape~\cite{book:reichl}.  However, quantum mechanics allows tunneling between KAM tori related by symmetry.  Dynamical tunneling was first discussed in the context of molecular physics by Davis and Heller~\cite{heller:dt}. 

Experiments have demonstrated that ensembles of ultracold atoms loaded into periodically modulated potentials provide a versatile test bed for studies of dynamical tunneling and more generally quantum chaos \cite{Moore_atomic_Rotor_PhysRevLett,exp:hensinger,exp:raizen,Hensinger_2003_JPB,otago:rotor1,otago:rotor2,Chabe_3dkickedrotor_PhysRevLett,Shrestha:DT_BECexp_PRE,Gadway_tworotors_PhysRevLett,Cheiney_matterwavemodul_PRA,Manai_2dkickedrotor_PhysRevLett,Sarkar_Rapol_kicked_rotor_PRL,Maurya_Rapol_kicked_rotor_PhysRevE,Cao_interactiondriven_dynlocbreakdown_NatPhys}. 
In particular, experiments allow the probing of quantum dynamical tunneling between symmetry-related islands of regular motion in phase space. This has been well-studied both experimentally~\cite{exp:hensinger,exp:raizen,chaudhury:kickedtop,Shrestha:DT_BECexp_PRE,Cheiney_matterwavemodul_PRA,Hainaut_exp_ideal_Floquet,Tomkovic_Poincare_Birkhoff_PRA,Dadras_quantwalk_PhysRevLett} and theoretically~\cite{heller:dt,mouchet:signatures,uterman:onsetofchaos, eltschka:resonances, hug:milburn, tomsovic:chaosassist, plata:classicalquantum, ballentine:tunneling,perez:tunneling, osovski:fingerprints, milburn:secondorderres,rebuzzini:nonlinear:dt,Sadgrove_phaseselect,Martinez_chaosassist_longrange_lattice_PhysRevLett,LIU201980,Schlagheck_enhancement_PhysRevLett,Kidd_BHdimer_modtunnel_PhysRevA,Eastman_controllingchaos_PhysRevA,Groiseau_spontem_PhysRevA}. Another arena for probing dynamical tunneling under controlled conditions is opto-mechanics~\cite{Jangid_DT_optomech_PhysRevA}.

Much of the theoretical work and the experiments with cold atoms on dynamical tunneling~\cite{exp:hensinger,exp:raizen} were performed in the single-particle regime where any interaction between particles can be reasonably neglected. In this case, the equations of motion are linear and the superposition principle applies.
However, many of the above experiments were performed using atomic gas Bose-Einstein condensates (BECs), where s-wave collisions between the atoms can be significant. In this situation BECs are better described by the Gross-Pitaevskii equation -- a nonlinear Schr\"odinger equation in which the nonlinear term captures these interactions. We have earlier shown that interactions unlock additional physics in dynamical tunneling such as self-trapping~\cite{wuester:trappingDT}.  Interactions will  have to be considered for experiments on dynamical tunneling with Bose-Einstein condensates on atom chips~\cite{martin:matthew:chip}.

Here we extend our earlier work on macroscopic nonlinear self-trapping in dynamical tunneling with Bose-Einstein condensates~\cite{wuester:trappingDT}.
Throughout we assume that the Gross-Pitaevskii equation is a suitable approximation to the dynamics.
The many-body theory of similar systems is considered e.g.~in Refs.~\cite{heimsoth:orbitaljosephson,heimsoth:effectivejosephson,Yulong_Grosspann_multiconf_FrontPhys,Satpathi_chaos_assit_manybody_PhysRevE,Hummel_JPA_2022}.
The system that we discuss is similar to that of the experiments~\cite{exp:hensinger,exp:raizen} which observed dynamical tunneling with an ensemble of ultra-cold atoms in a sinusoidal potential.
In the phase space of these atoms some regions classically form `islands of stability' amid a sea of chaos, which contain stable and regular periodic trajectories.
The symmetry of the system results in two such islands for which the classical trajectories have opposite momenta, with both oscillating in phase with the modulation of the potential.
Periodically sampling the atomic momentum for a classical trajectory on one of these islands would yield the same momentum each period, as the Kolmogorov-Arnold-Moser (KAM) theorem forbids a trajectory to cross from one island to another~\cite{book:reichl}.
The experiments, however, observe a slow oscillation of the periodically sampled atomic momentum between the islands which implies quantum mechanical tunneling between these classically disconnected regions of phase-space, i.e.\ dynamical tunneling.

It has been shown theoretically that, due to the presence of chaos, the period of dynamical tunneling exhibits an extremely irregular dependence on system parameters, such as the strength of the potential and of its modulation \cite{martin:matthew:chip,uterman:onsetofchaos,mouchet:resonances,mouchet:signatures}.
However the experimental investigation of these phenomena necessitates quite high atomic densities \cite{martin:matthew:chip}. This increases the strength of interactions and requires extensions of present theory which is applicable only in the \emph{linear regime}, i.e.\ at low atomic densities.
Nonlinearity has been shown to have an adverse effect on some transport phenomena: it can suppress resonance-enhanced Landau-Zener tunneling in optical lattices \cite{wimberger:wannier:stark}, and quantum transport in a kicked harmonic oscillator \cite{rebuzzini:nonlinear,rebuzzini:nonlinear2} and in a kicked rotor \cite{rebuzzini:nonlinear:acc, wimberger:kickedcond}.  However quantum transport in ratchet potentials can be enhanced~\cite{flach:ratchet} or suppressed \cite{heimsoth:orbitaljosephson}.
Dynamical tunneling of cold atoms may also be suppressed by strong nonlinearities \cite{rebuzzini:nonlinear:dt}. 

As we have shown in Ref.~\cite{wuester:trappingDT}, and further develop here, nonlinear interactions in a Bose-Einstein condensate, characterised by strength $U$, begin to suppress dynamical tunneling above a critical interaction strength, i.e.\ when $U>\sub{U}{crit}$.
This tunneling suppression is due to macroscopic quantum self-trapping (MQST) in close analogy to the suppression of BEC tunneling between two weakly-coupled potential wells that form a bosonic Josephson junction  \cite{smerzi:mqst,oberthaler:exp1,oberthaler:exp2,joel:MQST}. Here we provide further details about the techniques employed in \rref{wuester:trappingDT} and significantly extend our analysis to a broader range of parameter space that demonstrate additional features. In particular we map out the critical interaction strengths for a wide range of parameters which will be crucial to enable experimental explorations. We also highlight dynamical instabilities of selected Floquet states and propose protocols for the initialisation of an experiment in a Floquet state.

On the technical side these results are obtained by finding nonlinear Floquet states \cite{holthaus:NLfloq,flach:ratchet}, i.e.~condensate states that are invariant up to a phase under evolution with the nonlinear Gross-Pitaevskii equation (GPE) over one modulation period.
Evolving superpositions of these over many modulation periods
 shows clean dynamical tunneling despite the nonlinearity of the GPE.
Increasing the interaction strength $U$ in these simulations allows us to numerically find $\sub{U}{crit}$.
We then show that a two-mode model based on Floquet tunneling states explains self-trapping and accurately forecasts the critical interaction strength, $\sub{U}{crit}$, where it occurs.
The forecast is possible based just on the system parameters and the \emph{linear} Floquet states, i.e.\ the wave functions that reform up to a phase after one modulation period in the absence of 
interactions.
Finally, we show simulations of nonlinear dynamical tunneling with experimentally realizable initial states that are approximations to the exact Floquet states.
We then find dynamical tunneling periods that are slightly offset from the expected value and analyze this deviation in more detail.

We define our model and present the required linear and nonlinear Floquet techniques in \sref{modulations}.
Then we investigate the effect of nonlinearity on dynamical tunneling using nonlinear Floquet states (\sref{shutdown}).
In \sref{twomm} we present the two-mode model and the expression for the critical self-trapping nonlinearity, which are then compared with the full Gross-Pitaevskii solutions.
A parameter-space survey is the subject of \sref{parametersurvey}. \sref{imperfect} reports on simulations using an experimentally realizable initial state before we conclude in \sref{conclusions}.
\aref{atomchip} derives the atom chip potential, while \aref{nonlinear_numerics} outlines the numerical procedure we use to find nonlinear Floquet states. In \aref{twomm_appendix} we present the derivation of the two-mode model applied in our analysis in detail.

%%%%%%%%%%%%%%%%%%%%%%%%%%%%%%%%
%%%%%%%%%%%%%%%%%%%%%%%%%%%%%%%%
%%%%%%%%%%%%%%%%%%%%%%%%%%%%%%%%
\section{Condensates in a periodically modulated potential}
\label{modulations}

\begin{figure}
\centering
\epsfig{file={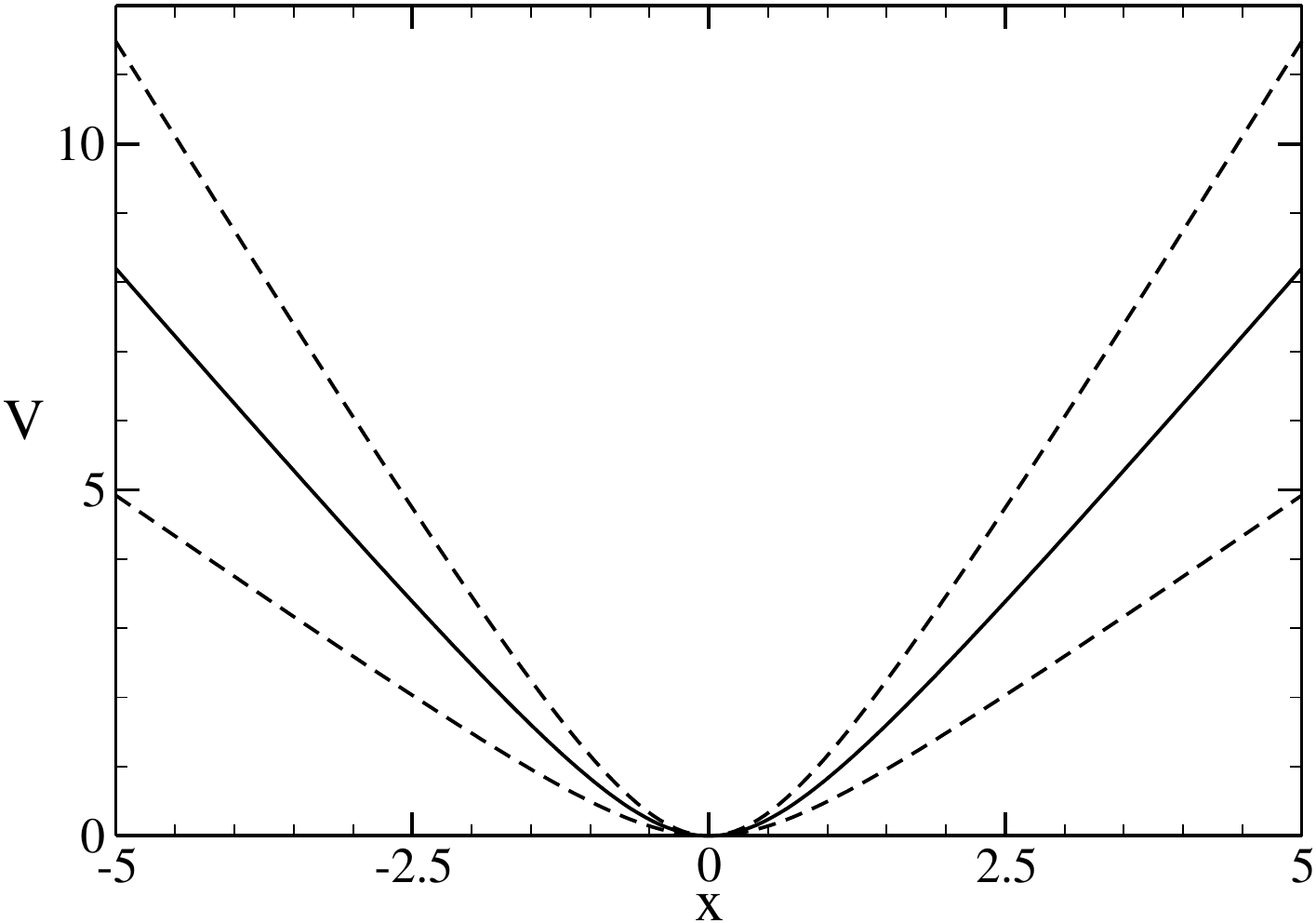},width=0.8\columnwidth} 
\caption{Anharmonic potential in the radial direction of an atom-chip trap, \eref{dimlesspotential}. The solid line indicates the mean potential and the dashed lines its largest modulation for the parameters $\kappa=2$, $\epsilon=0.4$.
\label{potential}}
\end{figure}

Atoms in a periodically-driven one-dimensional (1D) potential can only follow classically chaotic trajectories when the trap is anharmonic.
While the experiments \cite{exp:hensinger,exp:raizen} realized such a trap using a sinusoidal optical lattice potential, an alternative option that we investigate here is the radial potential typical for traps implemented with atom-chips. These are harmonic near the centre and become linear at large distances from the trap minimum as shown in \fref{potential}. This shape is dictated by the magnetic field strength profile near its local minimum~\cite{martin:matthew:chip}.

In this article we consider a Bose-Einstein condensate placed into a periodically modulated potential of this kind. This choice is largely motivated by practical considerations: As opposed to the sinusoidal potential, the chip potential keeps the condensate localized for a wide range of modulation parameters. We expect our main results --- the occurrence of dynamical tunneling despite nonlinear atomic self-interaction and of macroscopic self-trapping --- will also be realised in any other anharmonic potential. 

The dimensionless radial atom-chip potential is 
\begin{align}
V(x,t)&=\kappa \big[1 + \epsilon \cos(t) \big]\big[\sqrt{1+ x^{2}} - 1\big],
\label{dimlesspotential}
\end{align}
where $\kappa$ denotes the overall potential strength and $\varepsilon$ represents the magnitude of the modulation. Dashed lines in \fref{potential} illustrate the amplitude of this modulation for typical parameters.

When a nonlinear interaction term is included, and mean-field theory applied, the dynamics of the system follows the Gross-Pitaevskii equation:
\begin{align}
	i\kbar \frac{\partial}{\partial t} \psi(x,t)& =\left[ -\frac{\kbar^{2}}{2}\frac{\partial^{2}}{\partial x^{2}} + V(x,t) + U |\psi|^{2} \right]\psi(x,t),
\label{standardgpe}
\end{align}
where $U$ is the interaction strength parameter and $\kbar$ the effective Planck's constant. The latter arises through our choice of length and time scales as discussed in detail in Ref.~\cite{martin:matthew:chip}, where we also derive Eq.~\bref{standardgpe}. 

For a brief summary of the derivation from \rref{martin:matthew:chip} and the definition of the dimensionless parameters $\kappa$, $\kbar$ and $U$ in terms of the dimensional parameters of the atom chip problem, see \aref{atomchip}.
For the remainder of the main article we will work with the dimensionless variables.

% %
%%

%%%%%%%%%%%%%%\%%%%%%%%%%%%%%%%%%
%%%%%%%%%%%%%%%%%%%%%%%%%%%%%%%%
\subsection{Chaos and Floquet theory\label{floquet}}

\begin{figure}[ht]
\centering
\epsfig{file={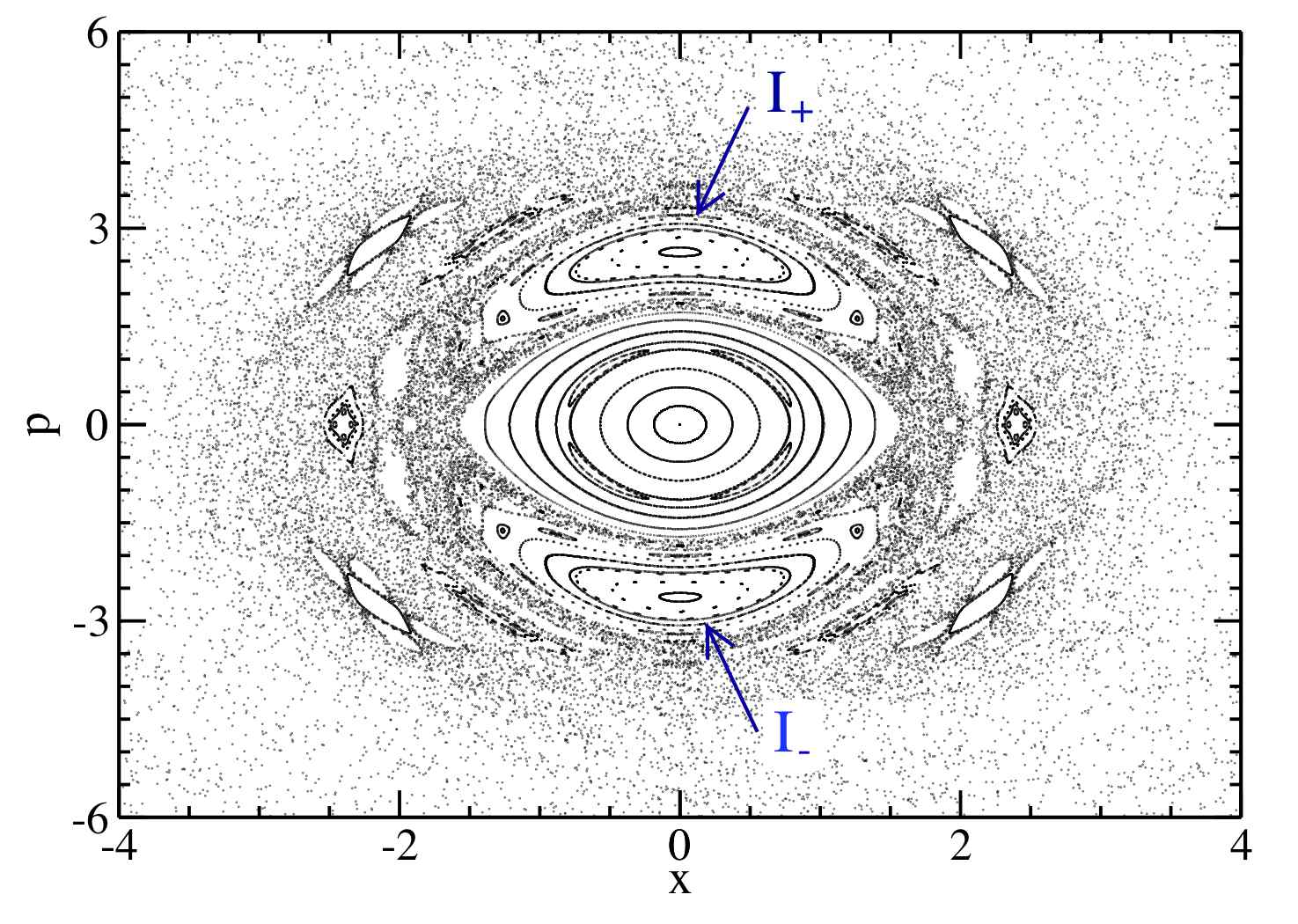},width=\columnwidth} 
\caption{Stroboscopic phase space for $\kappa=2$, $\epsilon=0.4$, showing islands of regular motion separated and surrounded by a region of chaos. Of primary concern for this paper are the two period-one fixed points located symmetrically on the $x=0$ axis, labelled $I_+$ and $I_-$.
\label{poincare_example}}
\end{figure}

The classical Hamiltonian for canonical coordinates ($p$, $x$) that corresponds to \eref{standardgpe} for $U=0$ is:
\begin{align}
H&=\frac{p^{2}}{2} + \kappa  \big[1 + \epsilon \cos(t) \big] \big[\sqrt{1+ x^{2}} -1 \big].
\label{class_hamil}
\end{align}
This Hamiltonian system can exhibit a mixed phase-space with chaotic and regular regions depending on the modulation parameters $\kappa$, $\epsilon$. For the case $\kappa=2$, $\epsilon=0.4$, we show a stroboscopic Poincar{\'e} section of phase space in \fref{poincare_example}.
For this we plot phase-space coordinates for a large range of initial conditions at fixed times $t=2\pi n$ for $n\in \mathbb{N}$. We point out a couple of key features: There are two large period-one islands of regular motion that are labelled $I_{\pm}$ in \fref{poincare_example}. They are
 symmetrically placed at $p=\pm p_{0}$, $x=0$ for some $p_{0}$. For trajectories located in these islands, atoms move in phase with the modulation of the potential. The islands are surrounded and separated by a sea of chaotic motion. Beyond these islands many higher order nonlinear resonances are visible \cite{book:reichl}.

For $\epsilon=0$ the Hamiltonian system (\ref{class_hamil}) is integrable and its phase-space hence regular. According to the KAM~theorem \cite{book:reichl}, we expect that regular regions persist in phase-space even if integrability is slightly broken, but become increasingly destroyed as $\epsilon$ is increased. The KAM~theorem also states that a classical trajectory from within one of the islands visible in \fref{poincare_example} may never cross into the other island.

Quantum mechanically this is no longer true. Consider the Schr\"odinger equation given by \eref{standardgpe} for $U=0$, i.e.\ the linear case with no interactions. Due to the time-dependence of the potential the usual description in terms of eigenstates of the Hamiltonian is not applicable. We can however resort to Floquet theory \cite{book:reichl}. Floquet's theorem states, that for a time-periodic Hamiltonian, $\hat{H}(t+T)=\hat{H}(t)$, a basis of the Hilbert space that solves \eref{standardgpe} can be found at all times in the form 
\begin{align}
|\chi_{i}(t)\rangle&=\exp{(-i\qe_{i} t/\kbar)} |u_{i}(t)\rangle,
\label{floquet_theorem}
\end{align}
where $\qe_{n}\in\mathbb{R}$ and $|u_{n}(T+t)\rangle=|u_{n}(t)\rangle$. We use ``$\chi$'' to denote the solution to \eref{standardgpe} in the linear $U=0$ case, reserving ``$\psi$'' for the nonlinear solutions found later. Although $|\chi_n(t)\rangle$ denotes a physical state, it is generally easier to first find the so-called Floquet state $|u_n(t)\rangle$.
This is because one can define a modified Hamiltonian $\hat{H}'(t)=\hat{H}(t) -i\kbar \partial/\partial t$ such that the Floquet states are its eigenstates:
\begin{align}
\hat{H}'(t) |u_{n}(t)\rangle = \qe_{n} |u_{n}(t)\rangle,
\label{Floquet_eqn}
\end{align}
where the eigenvalue $\qe_n$ is called the \emph{quasienergy}.

Equation~(\ref{floquet_theorem}) together with the periodicity of $u$ implies that $|\chi_{n}(t)\rangle$ is reformed after one period up to a unitary phase factor. This allows one to obtain $|u(x,t=0)\rangle$ as an eigenvector of the unitary time evolution operator over one period $\hat{U}(t+T,t)$ (which we distinguish by the operator hat from the dimensionless interaction strength $U$), with eigenvalue $\xi_{n}=\exp{(-i \qe_{n}T/\kbar)}$. To construct $\hat{U}(t+T,t)$, one can choose a basis of the Hilbert space and evolve each basis element over one modulation period according to \eref{standardgpe} with $U=0$. In practice we choose the basis spanned by symmetric and anti-symmetric discrete position space eigenstates which automatically decouples the even and odd subspaces.

\begin{figure}[ht]
\centering
\includegraphics[width=\columnwidth]{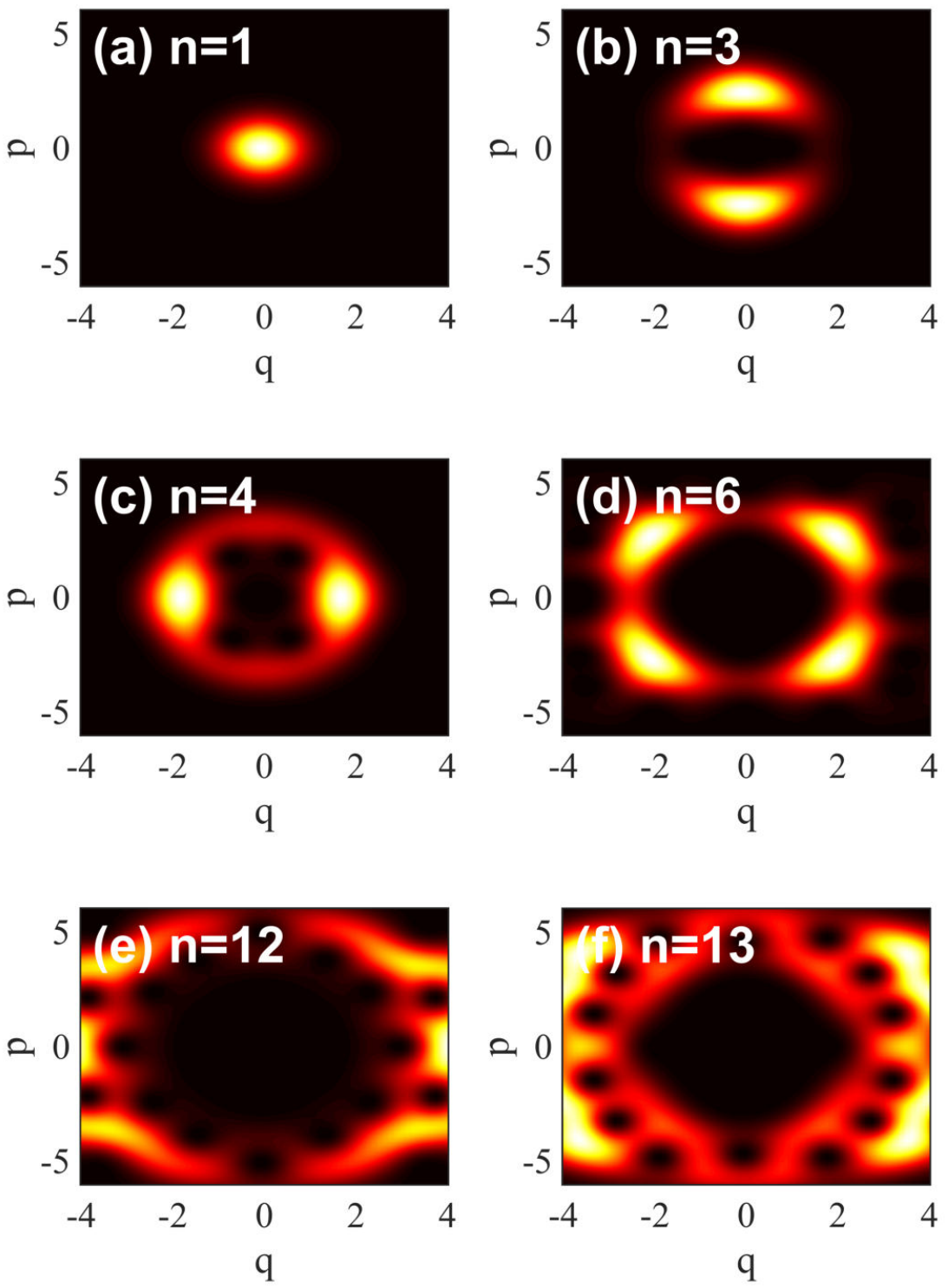} 
\caption{Selected Husimi functions, \eref{husimi_defn}, of even Floquet states $u_n$ for $\kappa$, $\epsilon$ as in \fref{poincare_example} and $\kbar=0.5$ (black $Q=0$, bright $Q$ large). Comparison with \fref{poincare_example} shows the correspondence between the distribution of Floquet states in phase-space and the classical distribution of chaotic and regular regions. The index $n$ labels the Floquet states, with energy at $t=0$ monotonically increasing for higher $n$. (a) regular region, (b) tunneling state, (c,d) further nonlinear resonances, (e,f) chaotic region.
\label{quantum_chaos}}
\end{figure}

In order to relate the linear Floquet states to the classical phase space (\fref{poincare_example}) one can make use of the Husimi (or $Q$) function of a quantum state $\Psi$ defined by:
\begin{align}
Q(p,x)[\Psi]&=\frac{1}{2\pi\kbar}\left|\langle \alpha | \Psi \rangle  \right|^{2},
\label{husimi_defn}
\end{align}
where $| \alpha\rangle$ is a coherent state centered on momentum $p$ and position $x$. The function $Q(p,x)[\Psi]$ gives information about the spread of $|\Psi\rangle$ in phase space due to the Heisenberg uncertainty. In \fref{quantum_chaos} we show $Q$-functions for various Floquet states with $\kbar=0.5$ that can be compared to the corresponding classical phase-space in \fref{poincare_example}. As expected \cite{book:reichl}, a remarkable correspondence is visible between the distribution of Floquet states over phase-space and the classical division into chaotic and regular regions.

One finds that the period-one islands of regular motion arise in the Floquet spectrum as a pair of states localized on \emph{both} islands, compare \frefp{quantum_chaos}{b}. These states are even (odd) under the transformation $p\rightarrow-p$ and have slightly different quasienergies $\qe_{n}$. In the following we will label these states $|u_{0}(t)\rangle$ (even) and $|u_{1}(t)\rangle$ (odd) and without atomic interactions, for $U=0$, we will refer to these particular Floquet states as \emph{linear tunneling states}. Floquet states have to arrange themselves with a definite parity under $p\rightarrow-p$, since the operator $\hat{H}'(t)$ is symmetric under this transformation.

An atomic wavepacket that is initially localized on just one of the period-one islands of \fref{poincare_example} must be represented as linear combination of the odd and even tunneling states: $|\chi_{\pm}(0)\rangle=[|u_{0}(0)\rangle \pm i |u_{1}(0)\rangle]/\sqrt{2}=[|\chi_{0}(0)\rangle \pm i |\chi_{1}(0)\rangle]/\sqrt{2}$, where the upper sign locates the atoms on the upper island.
The phase-factor $i$ results from our choice of phase of the potential modulation, as $|\chi_+(t=0)\rangle$ describes a particle with nonzero velocity and hence non-vanishing phase gradient. 
In the linear case it is easy to see that dynamical tunneling will occur. 
Using \eref{floquet_theorem} for $t=nT$, where $n\in\mathbb{N}$, we find the time evolution of $|\chi_{\pm}(t)\rangle$:
\begin{align}
|\chi_\pm(n T)\rangle&=e^{-i \qe_{0} n T/\kbar}  \CR
&\times \left[ |u_{0}(0)\rangle \pm i e^{i (\qe_{0}-\qe_{1}) n T/\kbar} |u_{1}(0)\rangle \right].
\label{lineartunneling}
\end{align}
The periodic change of sign of the second term causes transitions between $|\chi_\pm\rangle$ (see above) and thus quantum tunneling. 

The quasienergy splitting of the odd and even tunneling states determines the \emph{linear} period of dynamical tunneling (i.e.\ the tunneling period in the absence of interactions):
\begin{align}
\sub{T}{lin}&=\frac{2\pi\kbar}{|\qe_{0}-\qe_{1}|}.
\label{linear_period}
\end{align}
It has been found that the tunneling-period is strongly reduced if the tunneling Floquet states have significant overlap with the classically chaotic region \cite{uterman:onsetofchaos,Martinez_chaosassist_longrange_lattice_PhysRevLett}. This can be understood as tunneling assisted by classical diffusion accross the chaotic region. A further signature of this \emph{chaos-assisted tunneling} is a strong irregular sensitivity of $\sub{T}{lin}$ to the modulation parameters ($\kappa$, $\epsilon$) \cite{mouchet:signatures,martin:matthew:chip}.

As $\kbar$ is decreased, the quantum system becomes increasingly sensitive to small scale structures of the classical phase space. Examples for the latter are the higher-order resonance chains surrounding the fixed points in \fref{poincare_example}. 
When $\kbar$ is low enough for the quantum states to resolve these structures, order of magnitude changes of the linear tunneling period can result \cite{martin:matthew:chip,eltschka:resonances,Baecker2008,Loeck2010,Fritzsch_resonancechains_PhysRevE}.

%%%%%%%%%%%%%%%%%%%%%%%%%%%%%%%%
%%%%%%%%%%%%%%%%%%%%%%%%%%%%%%%%
\subsection{Nonlinear Floquet states}
\label{nonlinfloquet}

For nonzero interaction $U\neq0$ the theory becomes more complex because the nonlinearity means that we can no longer construct the time-evolution operator $\hat{U}$ from the evolution of a set of basis states.
In the following we outline how \emph{nonlinear} Floquet states can be constructed which were also used in \cite{holthaus:NLfloq,flach:ratchet}. We denote these states by $\phi$ to distinguish them from their linear counterparts $u$. They are solutions of
\begin{align}
\hat{H}'_{\rm nl}(t) \phi_{n} =\qenl_{n}(U)\phi_{n}, 
\label{nonlin_Floqeqn}
\end{align}
which is analogous to \eref{Floquet_eqn},
but with a nonlinear operator $\hat{H}'_{\rm nl}(t)$ defined by
\begin{align}
\hat{H}'_{\rm nl}(t) =  -\frac{\kbar^{2}}{2m}\frac{\partial^{2}}{\partial x^{2}} + V(x,t) + U | \phi_{n} |^{2} -i\kbar \frac{\partial}{\partial t},  
\label{nonlin_Floqoperator}
\end{align}%
that now includes a term describing particle interactions. We denote $\qenl_{n}(U)$ the nonlinear quasienergy for the interaction strength $U$, hence by definition, $\qe_n = \qenl_n(0)$.
We similarly change the labels for several variables to clearly distinguish the nonlinear case from the linear case as summarised in \tref{tab:variables}. 

\begin{table}
	\centering
	\begin{tabular}{|l|r|r|}\hline
		 & Linear & Nonlinear \\
		 & ($U=0$) & ($U> 0$) \\\hline 
		Quasienergies \rule{0pt}{11pt} & $ \lambda_n $ & $ E_n $ \\
		Physical state & $|\chi \rangle $ & $ |\psi \rangle $ \\
		Floquet state & $|u \rangle $ & $ |\phi \rangle $ \\
		Tunneling states & $|u_\pm \rangle $ & $ |\phi_\pm \rangle $ \\
		\hline
	\end{tabular}
	\caption{List of labels for different variables in the linear and nonlinear cases. }
	\label{tab:variables}
\end{table}

\begin{figure}
\centering
\epsfig{file={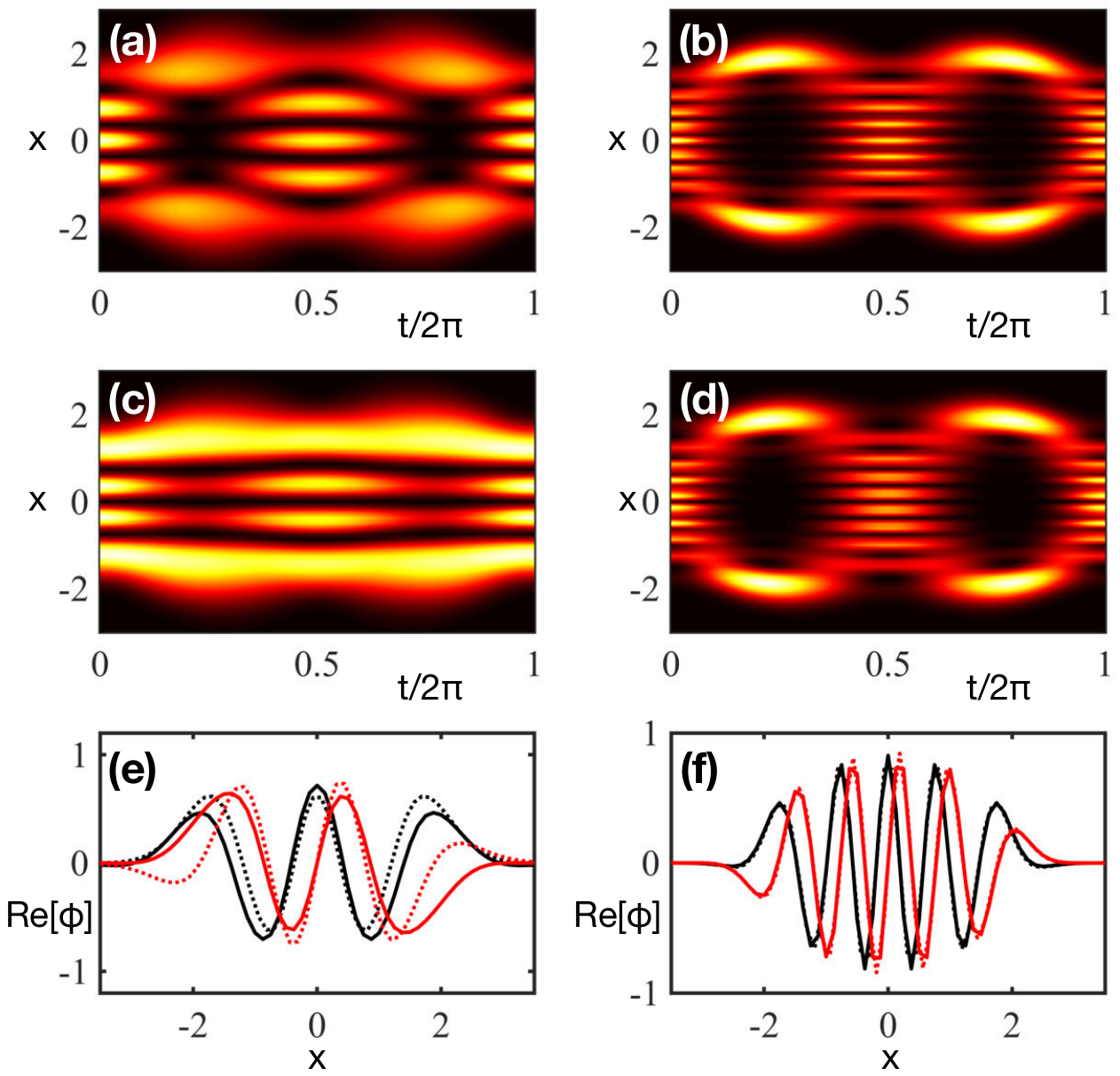},width=\columnwidth} 
\caption{ Examples of even and odd nonlinear Floquet tunneling states $\phi_{0,1}$. Physical parameters used are $\kappa=2$, $\epsilon=0.2$, $\kbar=0.5$, $U= 2$ for (a,c,e) and $\kappa=2$, $\epsilon=0.2$, $\kbar=0.25$, $U=0.003$ for (b,d,f). (a,b) Space-time densities of the even states and (c,d) of the odd ones. Black is zero, bright is high. (e,f) Comparison between the linear and nonlinear Floquet states, showing $\real{\phi}$ and $\real{u}$ at $t=\pi$. States are color coded: even-nonlinear (black-solid), even-linear (black-dotted), odd-nonlinear (red-solid), odd-linear (red-dotted). 
\label{example_nlfloquet}}
\end{figure}

The numerical procedure used to determine these solutions is much more involved than for the linear case involving an iterative optimisation scheme as outlined in \aref{nonlinear_numerics}.
Examples for the evolution of the odd and even nonlinear tunneling states over one period of the modulation are shown in \fref{example_nlfloquet}.

\begin{figure}[ht]
\centering
\epsfig{file={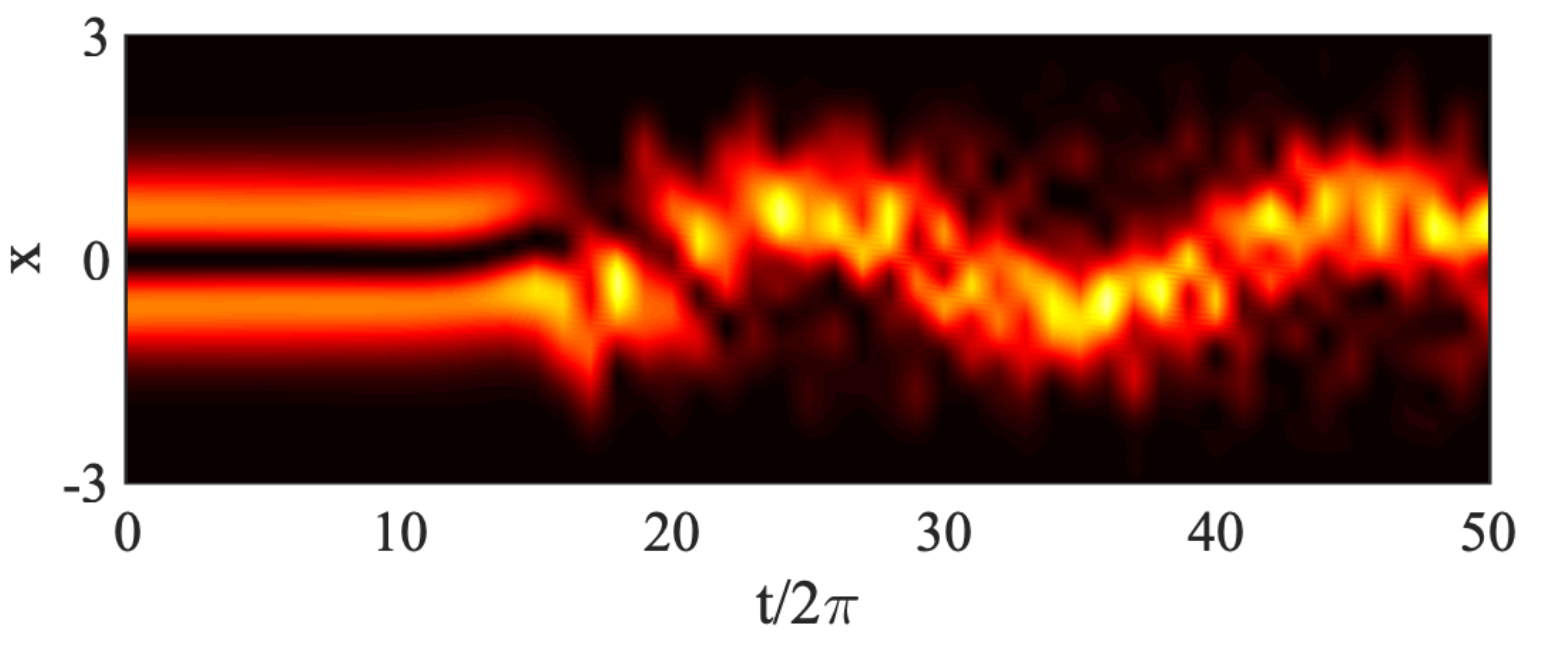},width=0.7\columnwidth} 
\caption{ Instability of the odd nonlinear Floquet state for a large nonlinearity $U=1.5$, $\kappa=1.4$, $\epsilon=0.2$, $\kbar=0.5$. Shown is the stroboscopically sampled momentum space density at $t=2\pi n$ for integer $n$, using the same colors as in \fref{example_nlfloquet}. After about 20 periods, we see the sudden onset of dynamical instability. The corresponding even nonlinear tunneling state is stable on this time scale. \
\label{instability}}
\end{figure}

For nonlinearities below the onset of self-trapping \cite{wuester:trappingDT} (considered in the next section) the nonlinear Floquet states are very similar to the linear ones. To benchmark our construction of nonlinear Floquet states, we hence sought states with much higher nonlinearities (up to $U\sim2$), as shown in \fref{example_nlfloquet}(a,c,e). At nonlinearities of this order the convergence of the algorithm becomes very slow. The states created show successful reformation for several periods as visible in \fref{instability}. However, after about 20 modulation periods, the figure shows a dynamical instability of the odd Floquet state, while the corresponding even state is stable for this time period. Clearly the stability properties of nonlinear Floquet states merit further systematic study but are beyond the scope of this paper. For the much smaller nonlinearity parameters used later in this article all nonlinear Floquet states were found to be stable.
%

%%%%%%%%%%%%%%%%%%%%%%%%%%%%%%%%
%%%%%%%%%%%%%%%%%%%%%%%%%%%%%%%%
%%%%%%%%%%%%%%%%%%%%%%%%%%%%%%%%
\section{Nonlinear dynamical tunneling}
\label{nonlinear}

We now use the nonlinear Floquet states to study dynamical tunneling in the presence of inter-atomic interactions. In \sref{shutdown} we give representative examples for dynamical tunneling and its arrest and in \sref{twomm} we explain our results with a two-mode model.

\begin{figure}[htb]
\centering
\epsfig{file={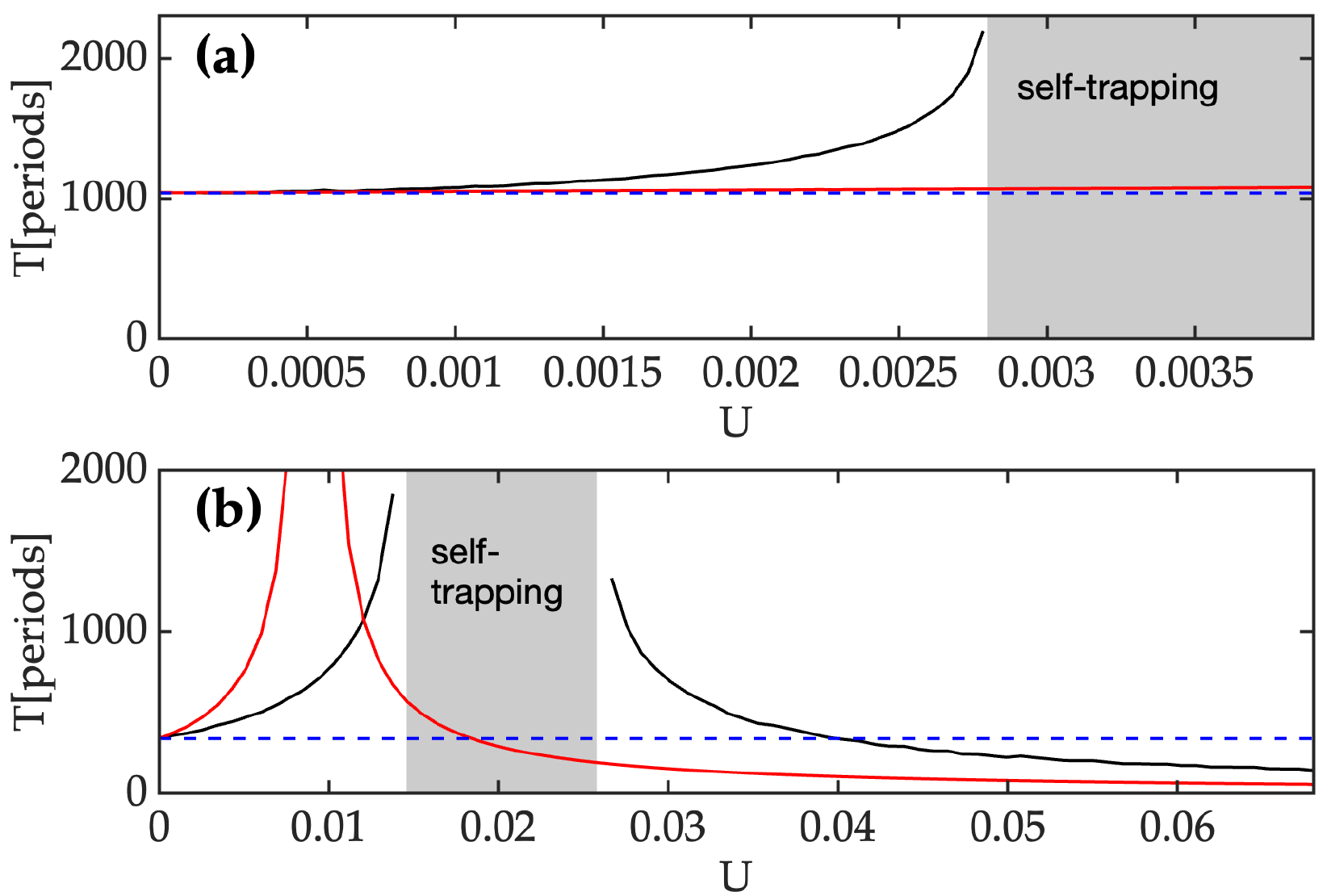},width=\columnwidth} 
\caption{Change of dynamical tunneling period with strength of nonlinearity. 
(a)  Estimate based on nonlinear quasi-energy differences in \eref{nonlinear_period} (red line) and actual period of tunneling in the nonlinear system, (black line), using the GPE \bref{standardgpe} for $\kbar=0.5$, $\kappa=2.3$ and $\epsilon=0.3$. We show the
linear tunneling period from \eref{linear_period} as a horizontal blue dashed line for comparison.
Within the gray shaded nonlinearity intervals, the GPE predicts self-trapping.
(b) The same, but for $\kbar=0.5$, $\kappa=1.3$ and $\epsilon=0.2$.
\label{periods_with_U}}
\end{figure}

One might expect that just as the tunneling period in the linear case follows \eref{linear_period}, the quantity:
\begin{align}
\sub{T}{nl}(U)&=\frac{2\pi\kbar}{|\qenl_{0}(U)-\qenl_{1}(U)|}.
\label{nonlinear_period}
\end{align}
might provide the nonlinear tunneling period.
However, it turns out that there are other dominant effects of the nonlinearity on the tunneling physics:
For small $U$, the value of $\sub{T}{nl}(U)$ increases slightly compared to $\sub{T}{lin}$, but we find that the actual tunneling period  deviates from this value as shown in \fref{periods_with_U}.
The reason for this is the violation of the superposition principle such that nonlinear dynamical tunneling is governed by a larger range of parameters than simply the nonlinear quasi-energy difference which enters \eref{nonlinear_period}, see \sref{twomm}.

%%%%%%%%%%%%%%%%%%%%%%%%%%%%%%%%
%%%%%%%%%%%%%%%%%%%%%%%%%%%%%%%%
\subsection{Nonlinear macroscopic self-trapping}
\label{shutdown}

\begin{figure}[t]
\centering
\epsfig{file={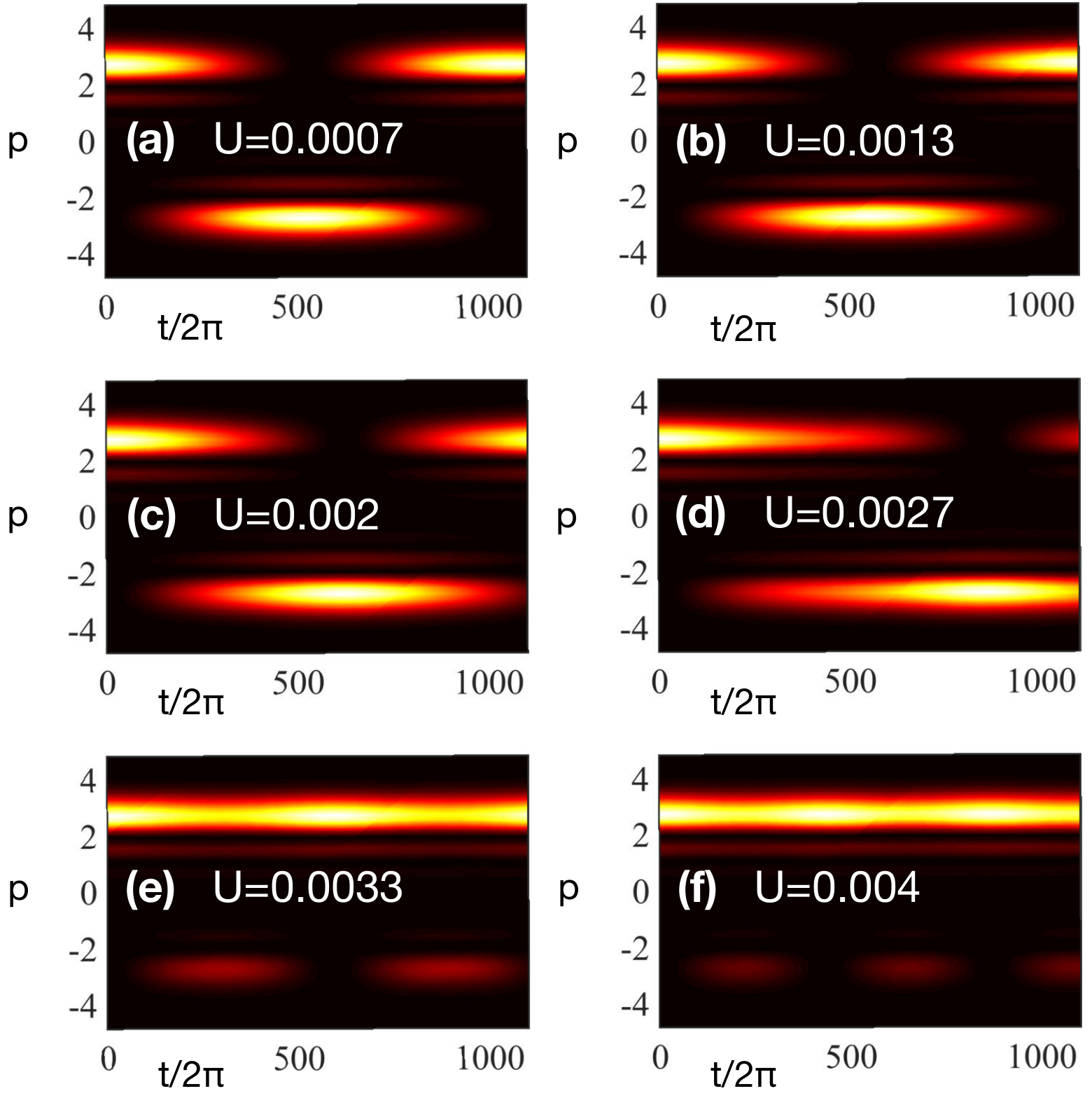},width=\columnwidth} 
\caption{ Dynamical tunneling in stroboscopically sampled momentum space, showing the momentum density (black zero, bright large), where $p=\kbar k$ with $\kbar=0.5$ and $\kappa=2.3$, $\epsilon=0.3$. Subsequent panels illustrate the onset of macroscopic quantum self-trapping as the nonlinearity is increased. For the chosen values of $\kbar$, $\kappa$, $\epsilon$, we have $\sub{T}{lin}=1038$ periods, \eref{linear_period}. If the tunneling period was determined by the \emph{nonlinear} quasienergies alone, i.e.~by $\sub{T}{nl}(U)$, it should increase by only $4\%$ from $\sub{T}{lin}$ over the range of $U$ shown. Clearly this is not the case, as the period is increased by nearly a factor of two between panels (a) and (d), and macroscopic self-trapping arises for (e) and (f).
\label{dyntunnel_momspace}}
\end{figure}

In this section we consider two specific scenarios that are representative of a large part of parameter space as will be shown in \sref{parametersurvey}. Let us first consider $\kappa=2.4$, $\epsilon=0.3$ and $\kbar=0.5$. From the linear tunneling states for this scenario we generate a series of nonlinear Floquet states up to $U=0.004$. 
Then we form the superposition localized on the upper resonance $\phi_{+}=\phi_{0}+i \phi_{1}$, and evolve it according to the GPE \bref{standardgpe} for $1100$ periods of the modulation. 
The simulations use an adaptive step-size Runge-Kutta integration method in the interaction picture, implemented in the high-level language XMDS \cite{xmds:paper,xmds:docu}.
Throughout the evolution, we stroboscopically sample the Fourier-spectrum $|\tilde{\psi}(k)|^2$ of the condensate wave function  after integer modulation periods. 
The momentum space picture thus obtained is shown in \fref{dyntunnel_momspace}.  For the chosen parameters, the classical resonance lies at $|p|=3.1$ on the momentum axis. As noted in \cite{hug:milburn} the quantum resonance can be slightly shifted in the presence of interactions.

It can be seen in \frefp{dyntunnel_momspace}{a} that the momentum space distribution is  essentially unchanged for many modulation periods. However on time scales still set by the quasienergy difference of the odd and even linear Floquet states, \eref{linear_period}, we observe a complete reversal of the sign of the condensate's momentum at the stroboscope instants $t=2\pi n$ with $n\in \mathbb{N}$. This is the signature of dynamical tunneling.  As the strength of the nonlinearity $U$ is increased we find that the period increases and for $U> 0.0027 $ complete momentum reversal no longer takes place. In this case the population becomes trapped in phase-space. This phenomenon is analogous to the arrest of inter-well tunneling in a bosonic Josephson junction due to strong nonlinearities, referred to as macroscopic quantum self-trapping~\cite{smerzi:mqst,oberthaler:exp1}. 
Macroscopic self-trapping effects on an even larger scale have been found for BECs in optical lattices \cite{anker:trapping,tristram:trapping}.
 
To check the accuracy of our nonlinear Floquet states and confirm that the tunneling effect is not due to numerical instability, we performed simulations like those in \fref{dyntunnel_momspace} but starting from the odd or even Floquet eigenstates rather than the tunneling states.
As expected, in these cases the stroboscopic wave function does not change shape on the time scale shown, because they are eigenstates of the one-period time-evolution operator.

\begin{figure}[ht]
\centering
\epsfig{file={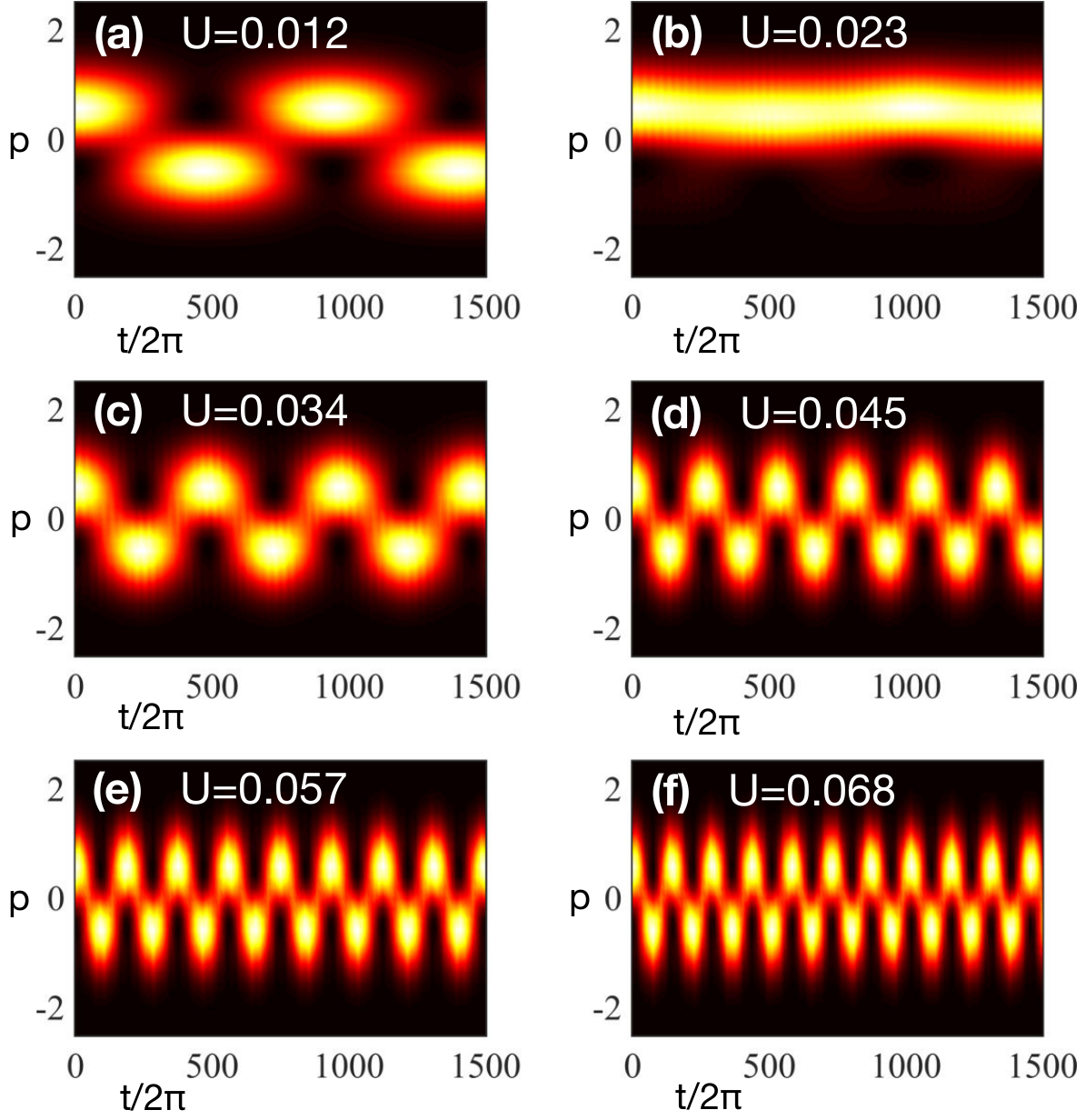},width=\columnwidth} 
\caption{ Dynamical tunneling in stroboscopically sampled momentum space $|\tilde{\Psi}(k,t=nT)|^2$, showing the density (black zero, bright large), where $p=\kbar k$ with $\kbar=0.5$ and $\kappa=1.3$, $\epsilon=0.2$. In contrast to the case shown in \fref{dyntunnel_momspace}, dynamical tunneling is only suppressed for a finite range of $U$ and reappears as the interaction is further increased. Here the dynamical tunneling periods are strongly affected by the nonlinearity. They also deviate from those expected from the quasienergy difference of the Floquet states. A more detailed comparison of tunneling periods and quasienergy differences can be found in \cite{wuester:trappingDT}.
\label{dyntunnel_momspace2}}
\end{figure}
We found that usually the use of nonlinear Floquet states is not crucial to see dynamical tunneling and MQST in simulations with nonzero interaction: Simply starting with the linear tunneling state still yields tunneling with the expected period, as well as nonlinear self-trapping for $U>\sub{U}{crit}$. This is consistent with the fact that for these parameters the nonlinear states undergo only a small change of shape compared to the linear ones. 

For the scenario in \fref{dyntunnel_momspace} the wavepacket remains trapped in stroboscopic phase space as $U$ is further increased. We also find cases where trapping occurs for isolated ranges of $U$, and disappears as $U$ is increased further, as shown in \fref{dyntunnel_momspace2}. An analogous effect in the macroscopic self-trapping of double well tunneling has not been reported but should exist for certain parameter regimes.  This is studied in \rref{cree:DThamiltonian} where the current authors completely characterize the dynamical regimes of quantum tunnelling in two-mode mode models of the Gross-Pitaevskii equation with repulsive contact interactions.

%%%%%%%%%%%%%%%%%%%%%%%%%%%%%%%%
%%%%%%%%%%%%%%%%%%%%%%%%%%%%%%%%
\subsection{Two-mode theory}
\label{twomm}

To interpret the results of the preceding section we present a simple two-mode model (TMM) for dynamical tunneling of cold atoms. We summarize the main points here and leave the detailed derivation to \aref{twomm_appendix}.

Let us assume that for a given time-periodic system we have found the odd and even \emph{nonlinear} Floquet states centered on the islands of regular motion. As previously we denote these states by $\phi_{0}$ (even) and $\phi_{1}$ (odd). Also as before, the modes localised on the upper and lower islands are found by superposition $\phi_{+}=\frac{1}{\sqrt{2}}(\phi_{0}+i \phi_{1})$ (upper) and $\phi_{-}=\frac{1}{\sqrt{2}}(\phi_{0}-i \phi_{1})$ (lower). Due to the orthogonality of $\phi_{0}$ and $\phi_{1}$ the linear combinations $\phi_{\pm}$ are also orthogonal. We know that the functions $\phi_{n}$ $n\in{0,1}$ solve the nonlinear Floquet equation:
\begin{align}
\left(\hat{H}_{0}(x,t) + U | \phi_n(x,t)|^{2} -i\kbar \frac{\partial }{\partial t}\right)\phi_n(x,t)&=\qenl_{n}\phi_n(x,t).
\label{NLFloquet_eqn}
\end{align}
In this expression and in the following, the argument $U$ of the nonlinear quasienergies $\qenl_{n}(U)$ is suppressed.
Let us now assume that the dynamics of the time-dependent solution of the GPE \bref{standardgpe} $\psi(x,t)$ is restricted to a linear combination of physical states $\psi_{\pm}(x,t)$.
Because these physical states are equal to the Floquet states up to a phase, \eref{floquet_theorem}, we can just as easily write $\psi$ in terms of the upper and lower island Floquet modes $\phi_{\pm}(x,t)$, i.e.:
\begin{align}
\psi(x,t)&=c_{+}(t)\phi_{+}(x,t) + c_{-}(t)\phi_{-}(x,t) .
\label{gpeexpansion}
\end{align}
From \bref{gpeexpansion}, it is possible to derive equations of motion for $c_{\pm}(t)$ from the GPE \bref{standardgpe}. One arrives at nonlinear equations of the form:
\begin{subequations}
\label{updowneqns_sketch}
\begin{align}
i\kbar \frac{\partial }{\partial t} c_{+}
&=\bar{\qenl}c_{+} + \frac{\Delta \qenl}{2}c_{-} + {\mathcal N}[c_{-},c_{-}^{*},c_{+},c_{+}^{*},t]\\
%%%
i\kbar \frac{\partial }{\partial t} c_{-}
&=\bar{\qenl}c_{-} + \frac{\Delta \qenl}{2}c_{+} + {\mathcal N}[c_{+},c_{+}^{*},c_{-},c_{-}^{*},t]
\end{align}
\end{subequations}
where $\bar{\qenl}=(\qenl_{0} + \qenl_{1})/2$ and $\Delta \qenl=\qenl_{0} - \qenl_{1}$. The full expressions for the nonlinear terms ${\mathcal N}$ are given in  \aref{twomm_appendix}, \eref{nonlinear_terms}. They are proportional to $U$, depending on overlap integrals of powers of the even and odd Floquet states:
\begin{subequations}
\label{couplings}
\begin{align}
U_{01}(t)&=U\int dx |\phi_{0}(x,t)|^{2}|\phi_{1}(x,t)|^{2},
\\
A_{01}(t)&=U\int dx \phi_{0}(x,t)^{2}\phi_{1}(x,t)^{*2},
\\
U_{00}(t)&=U\int dx |\phi_{0}(x,t)|^{4},
\\
U_{11}(t)&=U\int dx |\phi_{1}(x,t)|^{4}.
\end{align}
\end{subequations}
Had we used a more complete expansion instead of \eref{gpeexpansion}, the right hand sides of \eref{updowneqns_sketch} would contain terms coupling to further nonlinear states, whose coefficients are proportional to $U$ and the overlap between the tunneling states and the additional nonlinear states in question. Figure \ref{quantum_chaos} suggests that, at least for the linear case, the overlap between the tunneling states and other Floquet states is small. Since $U$ will be small as well one might expect that other nonlinear Floquet modes besides the tunneling states can be neglected, which is indeed what we usually find numerically. 

We consider two different methods to predict the shutdown of dynamical tunneling at a critical nonlinearity from \eref{updowneqns_sketch}. In the \emph{explicit} method we numerically extract the functional form of the coefficients in \eref{couplings} over one period from the nonlinear Floquet states obtained in \sref{nonlinfloquet}. Equation \bref{updowneqns_sketch} can then be solved numerically. Importantly, this has  reduced the problem from a 1D partial differential equation into two coupled ordinary differential equations. In the \emph{averaged} method, we additionally replace the coefficients by their average over one period, e.g.~$U_{ij}(t)\rightarrow \bar{U}_{ij}=\int_{0}^{T}U_{ij}(t)dt/T$. This is justified when the oscillations in the populations of the modes $\phi_{+}$ and $\phi_{-}$ take place on time scales much larger than a single period. This approximation was found to be well satisfied for most cases inspected. 

\begin{figure}[ht]
\centering
\epsfig{file={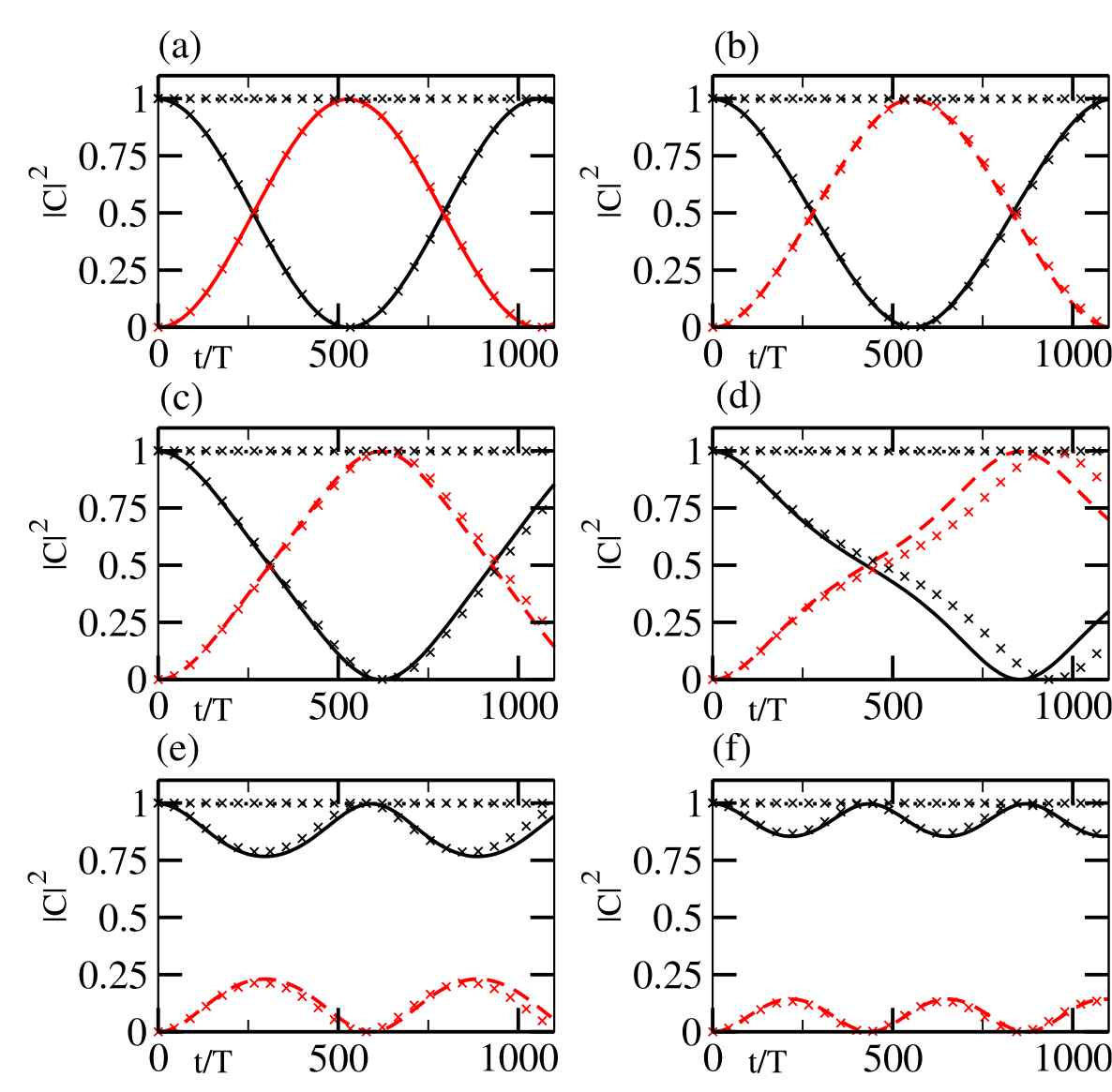},width=\columnwidth} 
\caption{ Comparison of GPE and two-mode model results for the coefficients of Floquet states centered on the upper and lower island of stability. The assignment of nonlinearities to the panels is as in \fref{dyntunnel_momspace}.  Lines indicate the GPE Floquet state populations $d_{\pm}(t)=\int dx \phi_{\pm}^{*}(x,t) \psi(x,t)$ and crosses their TMM counterparts $|c_{\pm}|^2$, see \eref{gpeexpansion}.  Black-solid: $|d_{+}|^{2}$, red-dashed: $|d_{-}|^{2}$, black-dotted: $\sub{n}{tot}\equiv |d_{+}|^{2} + |d_{-}|^{2}$, black crosses $|c_{+}|^{2}$ or $\sub{n}{tot,2}\equiv |c_{+}|^{2} + |c_{-}|^{2}\approx1$, red crosses: $|c_{-}|^{2}$. The coupling coefficients for the two-mode model, \eref{couplings}, have been determined from the appropriate nonlinear Floquet states using the explicit method discussed in the text. 
\label{twomode_coefficients}}
\end{figure}

\begin{figure}[ht]
\centering
\epsfig{file={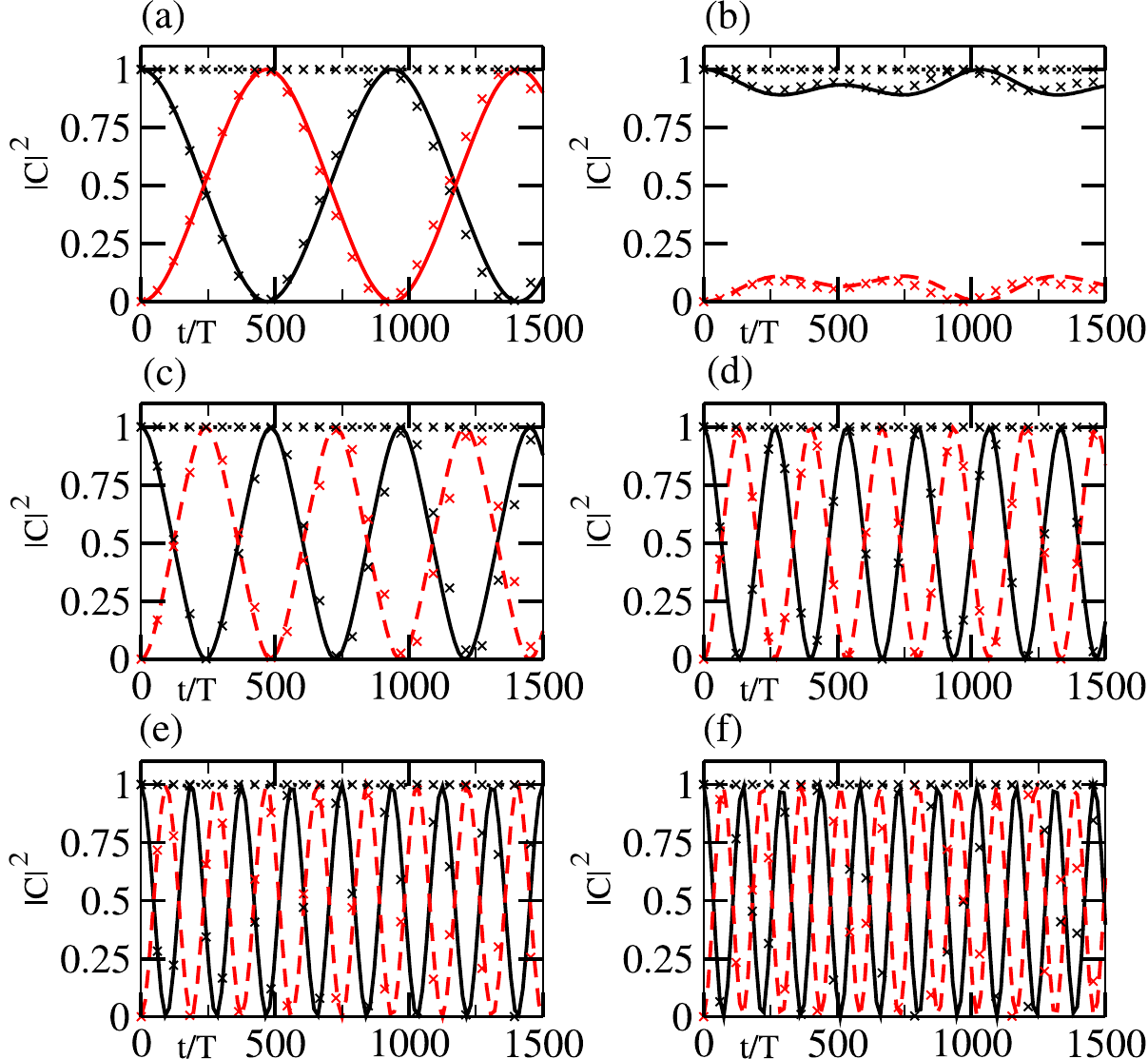},width=\columnwidth} 
\caption{ GPE vs.~two-mode model results for a scenario without complete trapping. Nonlinearities as in \fref{dyntunnel_momspace2}. Lines are coded as in \fref{twomode_coefficients}.
\label{twomode_coefficients2}}
\end{figure}

We validated the two-mode results by direct comparison with full solutions of the GPE. To this end we extract the coefficients $d_{+}(t)=\int dx \phi_{+}^{*}(x,t) \psi(x,t)$ and $d_{-}(t)=\int dx \phi_{-}^{*}(x,t) \psi(x,t)$ from the numerical data.
This makes use of the nonlinear Floquet states $\phi_{\pm}(x,t)$ as defined at the beginning of this section, and our solution $\psi(x,t)$ of \eref{standardgpe}.
These coefficients can be directly compared with coefficients $c_{\pm}$ computed using the two-mode model, \eref{updowneqns_sketch}, with either explicit or averaged coefficients as explained earlier. Figure \ref{twomode_coefficients} shows this comparison for the same case as \fref{dyntunnel_momspace}, and \fref{twomode_coefficients2} corresponds to \fref{dyntunnel_momspace2}. We see that the TMM quite accurately captures the complete solution of the GPE.

A key advantage of the averaged method is that it allows additional analytical insight since the coefficients no longer depend on time.
Following \cite{smerzi:mqst}, we can cast the TMM into a more illuminating form by reformulating it in terms of the population imbalance $z=N_{+}-N_{-}$, and relative phase $\varphi=\theta_{-}-\theta_{+}$, where $c_{\pm}=\sqrt{N_{\pm}}e^{i\theta_{\pm}}$ with $N_{\pm}, \theta_{\pm} \in \mathbb{R}$, see \aref{twomm_appendix}.
The resulting equations of motion for $z$ and $\varphi$ can be derived as Euler-Lagrange equations for canonical variables $z$, $\varphi$ from the effective classical Hamiltonian:
\begin{align}
\sub{H}{eff}&=\frac{1}{2} \Lambda z^{2} +\alpha \sqrt{1-z^{2} }\cos{(\varphi)} + \beta (1-z^{2})\cos{(2\varphi)},
\label{effhamiltonian}
\end{align}
with parameters
\begin{subequations}
\begin{align}
\Lambda&=\frac{\bar{U}_{00}+\bar{U}_{11} }{4} +\frac{3}{2} \real{\bar{A}_{01}}-\bar{U}_{01},
\\
\alpha&=\frac{\bar{U}_{00}-\bar{U}_{11} }{2} - \Delta \qenl, 
\\
\beta&=\frac{\bar{U}_{01}}{2} -\frac{\bar{U}_{00}+\bar{U}_{11} }{8} + \frac{ \real{\bar{A}_{01} } }{4}.
\end{align}
\end{subequations}
Equation \bref{effhamiltonian} is identical to the Hamiltonian of \rref{smerzi:mqst} for $\alpha=1$ and $\beta=0$. 
As shown in \cite{smerzi:mqst}, an effective conservative Hamiltonian allows us to determine initial conditions $(z(0),\varphi(0))$ for which full dynamical tunneling can never take place.  

Dynamical tunneling requires that the population imbalance $z$ changes from positive to negative, or vice versa, so the system must pass through some phase space point where $z=0$.
If the initial conditions have an effective energy value that is incompatible with $z=0$, the phase-space structure encoded by \eref{effhamiltonian} prevents a trajectory with such initial conditions from passing through $z=0$, hence the trajectory must correspond to self-trapped dynamics.
In our case the analysis is more complicated than in \rref{smerzi:mqst} because of the $\cos{(2\varphi)}$ term in the effective Hamiltonian. 

To use the above arguments to infer an analytical critical nonlinearity, we assume that initially $z(0)=1$, $\varphi(0)=0$, and demand $z(t^*)=0$ at some time $t^*$ if tunneling is to take place. 
As the classical effective Hamiltonian is time-independent, energy conservation applies and guarantees that $H(0)=H(t^*)$, which in this case implies: 
\begin{align}
\frac{1}{2}\Lambda&=\alpha \cos{(\varphi(t^*))} + \beta \cos{(2\varphi(t^*))}.
\label{maxcoupling1}
\end{align}
Thus a sufficient condition for self-trapping is that this equation have no solution for any $\varphi$.
We define $\sub{U}{c,full}$ as a critical nonlinearity above which this is the case.
Obtaining it analytically is in general difficult for reasons explained in \aref{twomm_appendix}. 

It becomes possible, hower, using a simple approximation that we found justified in most cases: We assume $|\Delta \qenl| \gg |\beta|$ and $|\Delta \qenl| \gg|(\bar{U}_{00}-\bar{U}_{11})/2|$. It can then be shown that for interaction strengths in excess of the critical value
\begin{align}
\sub{U}{crit}&=\frac{2 |\Delta \qenl|}{|\Lambda_{0}|},
\label{critical_coupling}
\end{align}
where $\Lambda_{0}=\Lambda/U$, \eref{maxcoupling1} does not have a solution and dynamical tunnelling undergoes macroscopic quantum self-trapping. 
For a detailed exploration of regions in the phase-space for dynamical tunneling encoded by \eref{effhamiltonian} without this approximation we refer to \rref{cree:DThamiltonian}.

Since $\Lambda_{0}$ no longer explicitly depends on $U$ we can define a corresponding quantity 
for the \emph{linear} Floquet states as well:
\begin{align}
\sub{\Lambda}{0,lin}=\frac{\sub{\bar{U}}{00,lin}+\sub{\bar{U}}{11,lin} }{4} +\frac{3}{2} \real{\sub{\bar{A}}{01,lin}}-\sub{\bar{U}}{01,lin},
\label{qelin}
\end{align}
 where
\begin{subequations}
\label{couplings_linear_approx}
\begin{align}
\sub{\bar{U}}{01,lin}(t)&=\frac{1}{T}\int dx\int dt\:\: |u_{0}(x,t)|^{2}|u_{1}(x,t)|^{2},
\\
\sub{\bar{A}}{01,lin}(t)&=\frac{1}{T}\int dx\int dt\:\:  u_{0}(x,t)^{2}u_{1}(x,t)^{*2},
\\
\sub{\bar{U}}{00,lin}(t)&=\frac{1}{T}\int dx \int dt \:\:|u_{0}(x,t)|^{4},
\\
\sub{\bar{U}}{11,lin}(t)&=\frac{1}{T}\int dx \int dt \:\:|u_{1}(x,t)|^{4}.
\end{align}
\end{subequations}
If we now assume that the parameters $|\Delta \qenl|$, $\Lambda_{0}$, in \eref{critical_coupling} do not change dramatically as the nonlinearity is increased we can estimate the critical nonlinearity for self-trapping simply from knowledge of the \emph{linear} Floquet states. That is, we define the \emph{linear forecast} of the critical nonlinearity
\begin{align}
\sub{U}{c,lin}&=\frac{2 |\Delta \qe|}{|\sub{\Lambda}{0,lin}|},
\label{critical_coupling_forecast}
\end{align}
where $\Delta \qe= \qe_1 -  \qe_0$ is the quasienergy splitting of linear tunnelling states from \eref{Floquet_eqn}. We show in the next section that \eref{critical_coupling_forecast} can indeed reliably predict the onset of trapping for a wide variety of modulation parameters.

To explain intermittent trapping as in \fref{dyntunnel_momspace2} and \fref{twomode_coefficients2}, note that \eref{maxcoupling1} 
does not always predict self-trapping in the limit $U\rightarrow \infty$. While the approximations invoked to find \eref{critical_coupling} might be appropriate for some intermediate range of $U$ and thus correctly yield the critical nonlinearity for the initial onset of self-trapping, those approximations can later become invalid as $U$ is increased further. In the limit $U\rightarrow \infty$ we can eventually neglect $\Delta E$ in comparison with all other terms, and then cancel a factor of $U$ from both sides of \eref{maxcoupling1}.
The latter now depends on $U$ only implicitly through the dependence of the nonlinear Floquet states $\phi_{0,1}(x,t)$, 
which therefore determines whether or not self-trapping persists to very large values of $U$.

For the cases shown here (and most other cases inspected) the two-mode model accurately reproduces the full dynamics, which is reasonable as we find that no significant population is transferred to modes beyond the odd and even nonlinear Floquet states. For the case in \fref{twomode_coefficients} the difference between the two-mode model calculations with averaged or explicit coefficients is not visible on the scale of the plot. The scenario in \fref{twomode_coefficients2} is however significantly better described by the explicit model, which is the comparison made.

In summary, we have devised a two-mode model  for dynamical tunnelling in Bose-Einstein condensates based on superpositions of the odd and even nonlinear Floquet states.
Despite the absence of the superposition principle, we find that this model can accurately describe dynamical tunnelling for a wide range of parameters. Under conditions where coupling coefficients in the two-mode model can be replaced by their time averages the model then allows analytical predictions of the critical non-linearity for the arrest of dynamical tunneling oscillations.

%%%%%%%%%%%%%%%%%%%%%%%%%%%%%%%%
%%%%%%%%%%%%%%%%%%%%%%%%%%%%%%%%
\section{Survey of critical interaction strengths}
\label{parametersurvey}
Here we extend our investigations of \sref{shutdown} to a broad range of modulation parameters $\kappa$, $\epsilon$ and effective Planck constants $\kbar$. We determine the critical nonlinearity for self trapping throughout parameter space and investigate how successfully we can use \eref{critical_coupling} to predict the critical nonlinearity for self-trapping from the linear Floquet states.

\begin{figure}[ht]
\centering
\epsfig{file={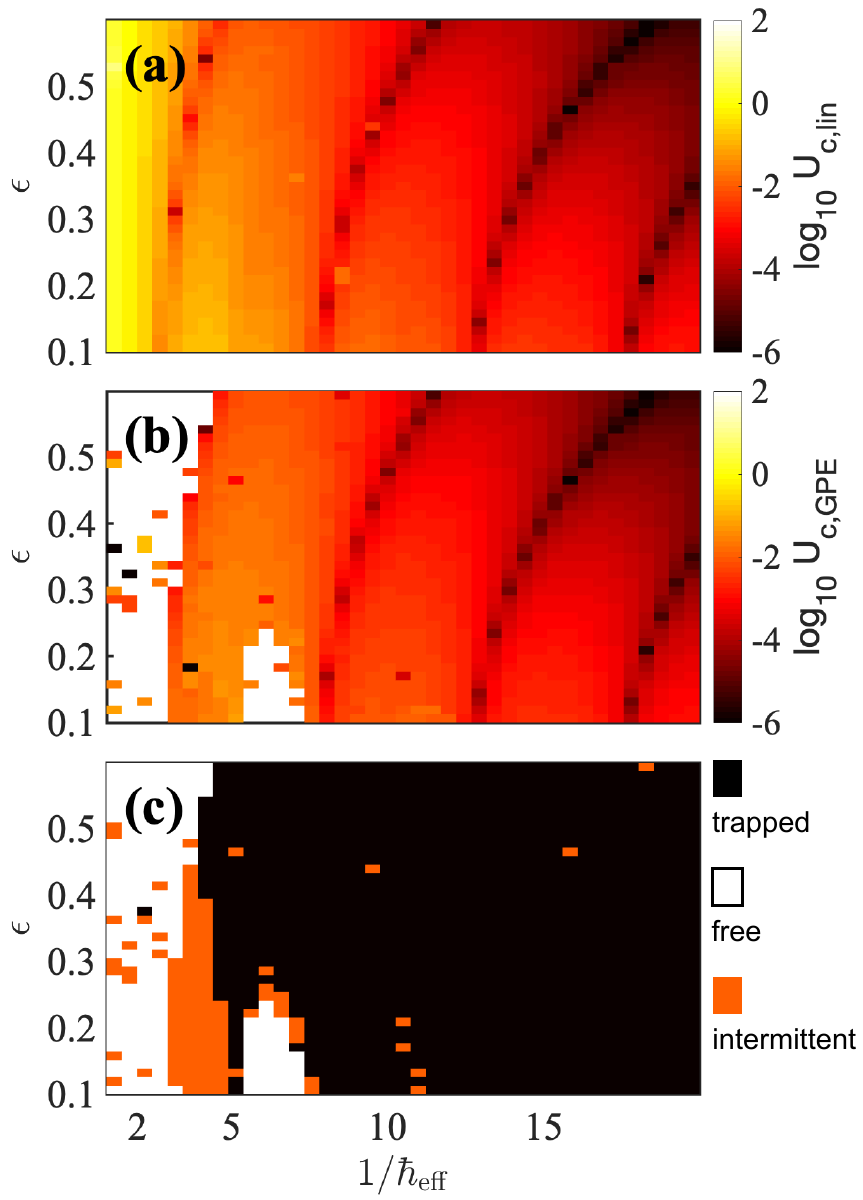},width=\columnwidth} 
\caption{Dependence of macroscopic quantum self-trapping features in dynamical tunnelling on $\kbar$ and $\epsilon$, for fixed $\kappa=1.2$ (c.f.~Fig.~3(a) of \rref{martin:matthew:chip}). (a) Linear forecast of critical nonlinearity, $\log_{10}(\sub{U}{c,lin})$, see \eref{critical_coupling_forecast}. (b) Actual critical nonlinearity $\log_{10}(\sub{U}{c,GPE})$, where $\sub{U}{c,GPE}$ was determined by solving the GPE \bref{standardgpe} for a range of nonlinearities $U$. The white areas indicate where no self-trapping occurs, i.e.~the tunnelling period is infinite. (c) Classification of trapping in the full GPE model. We distinguish three cases as discussed in the text: (i) free (white), trapped (black) and trapped only for an intermittent range of interaction strengths (orange).}
\label{2Dparamscan}
\end{figure}
\begin{figure}[ht]
\centering
\epsfig{file={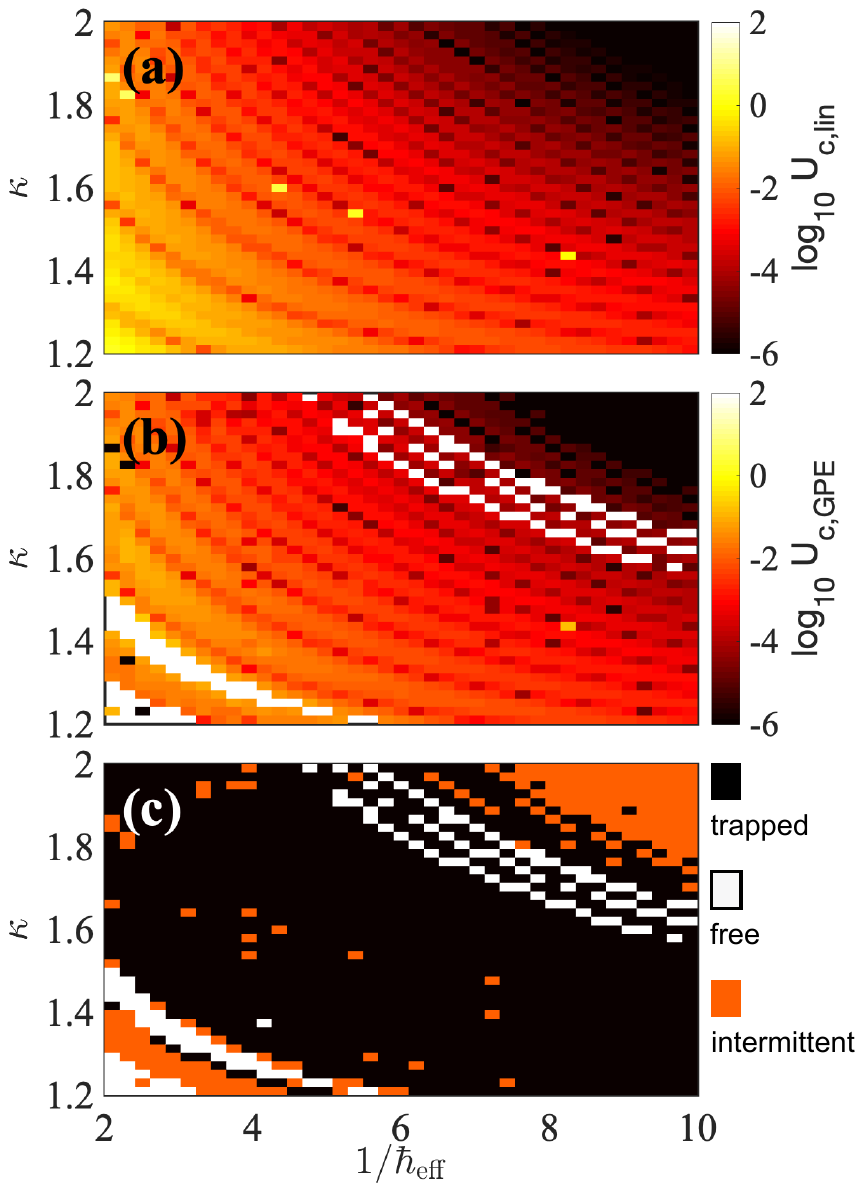},width=\columnwidth} 
\caption{Same as in \fref{2Dparamscan}, but showing the dependence on on $\kbar$ and $\kappa$ for fixed $\epsilon=0.2$. 
\label{2Dparamscan2}}
\end{figure}
We concentrate on a pair of two-dimensional slices of the 3D parameter space ($\kappa$, $\epsilon$, $\kbar$), sampling critical nonlinearities and self-trapping characteristics as a function of $\epsilon$ and $\kbar$ at fixed $\kappa=1.2$ (see \fref{2Dparamscan}) and as a function of $\kappa$ and $\kbar$ at fixed $\epsilon=0.2$ (see \fref{2Dparamscan2}). For the former parameter range, linear tunneling periods were known from \rref{martin:matthew:chip}.

For all data points (parameter sets) we used the following numerical protocol:
\begin{enumerate}
\item Generate linear Floquet states, see \sref{floquet}. Use these to estimate the critical interaction strength for self-trapping $\sub{U}{c,lin}$ from \eref{critical_coupling_forecast} (see \fref{2Dparamscan}(a)). 
\item Generate nonlinear Floquet states, see \sref{nonlinfloquet}, for a set of interaction strengths $U_j$ up to roughly $\sub{U}{trial}=2\times\sub{U}{c,lin}$. Use each of these to generate an
initial state $\psi(x,t=0)$ for the GPE \bref{standardgpe} that is centered on one of the islands of stability. \label{NLFstep}
\item Simulate dynamical tunneling oscillations for all $U_j$ with the GPE \bref{standardgpe}. Classify each $U_j$ as untrapped (trapped) in the GPE model if the mean momentum $\langle \hat{p} \rangle$  crosses (does not cross) zero.
Determine the true critical nonlinearity $\sub{U}{c,GPE}$ as the first trapped $U_j$ (see \fref{2Dparamscan}(b)).
\item Categorise each point (see \fref{2Dparamscan}(c)) as (i) ``free" if all $U_j$ are untrapped. (ii) ``trapped" if only all $U_j$ above a given critical $U$ are trapped, e.g.~as in \fref{dyntunnel_momspace}. (iii) ``intermittently trapped" if a continuous range of $U_j$ are trapped \cite{footnote:isolated_values}, but the highest investigated value is not, e.g.~as in \fref{dyntunnel_momspace2}.
\end{enumerate}
The results for $\sub{U}{c,lin}$ and $\sub{U}{c,GPE}$ are shown in \fref{2Dparamscan} and \fref{2Dparamscan2} together with the characterisation of trapping for the GPE. Note that the characterisation depends on $\sub{U}{trial}$ which we cannot choose arbitrarily large due to numerical constraints. It can be seen in \fref{2Dparamscan} and \fref{2Dparamscan2} that for most trapped cases the linear forecast of the critical nonlinearity gives a good idea of the true critical nonlinearity. This is particularly the case when $\sub{U}{c,lin}$ is small, such that nonlinear Floquet states do not deviate much from linear ones, and hence \eref{critical_coupling_forecast} is a very good approximation. For many cases where $\sub{U}{c,lin}$ is large enough such that Floquet-states with $U\approx \sub{U}{c,lin}$ are significantly modified compared to the linear ones, this results in the removal of trapping, or significantly increases the actual $\sub{U}{c,GPE}$ (which we can not distinguish in our numerics due to the finite $\sub{U}{trial}$). This occurs for example for the largest $\kbar$ values in \fref{2Dparamscan}.

The nonlinear two-mode model based on \eref{maxcoupling1} predicts these larger untrapped regions and the lobe of intermittently trapped parameters around $\kbar^{-1}\approx4$ and $\epsilon\approx2$ in \fref{2Dparamscan} quite well (not shown). However there are a few isolated data points throughout at which predictions differ from the GPE. Given that we detect some unstable nonlinear Floquet states as in \fref{instability} for the larger nonlinearities, we consider that overall the predictions of the two-mode model are highly reliable. 
However, this may be a sign that the self-trapping transition is robust to imperfections -- an idea that we will explore further in the next section.

The overall trend visible in all panels of \fref{2Dparamscan} and \fref{2Dparamscan2} is that critical interaction strengths $U_c$ decrease as $\sub{h}{eff}$ is reduced. 

%%%%%%%%%%%%%%%%%%%%%%%%%%%%%%%%
%%%%%%%%%%%%%%%%%%%%%%%%%%%%%%%%
%%%%%%%%%%%%%%%%%%%%%%%%%%%%%%%%
%%%%%%%%%%%%%%%%%%%%%%%%%%%%%%%%
\section{Experimental preparation of dynamical tunneling states}
\label{imperfect}
An important consideration is how our predictions can be realized in experiment. The Floquet states that undergo dynamical tunneling are not straightforward to precisely prepare in an experiment. In this section we propose a scheme to prepare good approximations to them, and then consider the impact of the resulting imperfect initial state $\psi(x,t=0)$ on the results presented so far. 

%%%%%%%%%%%%%%%%%%%%%%%%%%%%%%%%
%%%%%%%%%%%%%%%%%%%%%%%%%%%%%%%%
\subsection{Approximating Floquet states with Gaussians}
\label{floquetapprox}
In the experiments \cite{exp:hensinger,exp:raizen}, the atomic cloud was initially prepared in a state resembling the tunneling Floquet state simply by shifting the potential by an appropriate amount. Along similar lines, here we consider providing a momentum kick $p$ consistent with the islands of stability to a condensate wave function in a trap ground state, see also the description in \rref{martin:matthew:chip}.

\begin{figure}[ht]
\centering
\epsfig{file={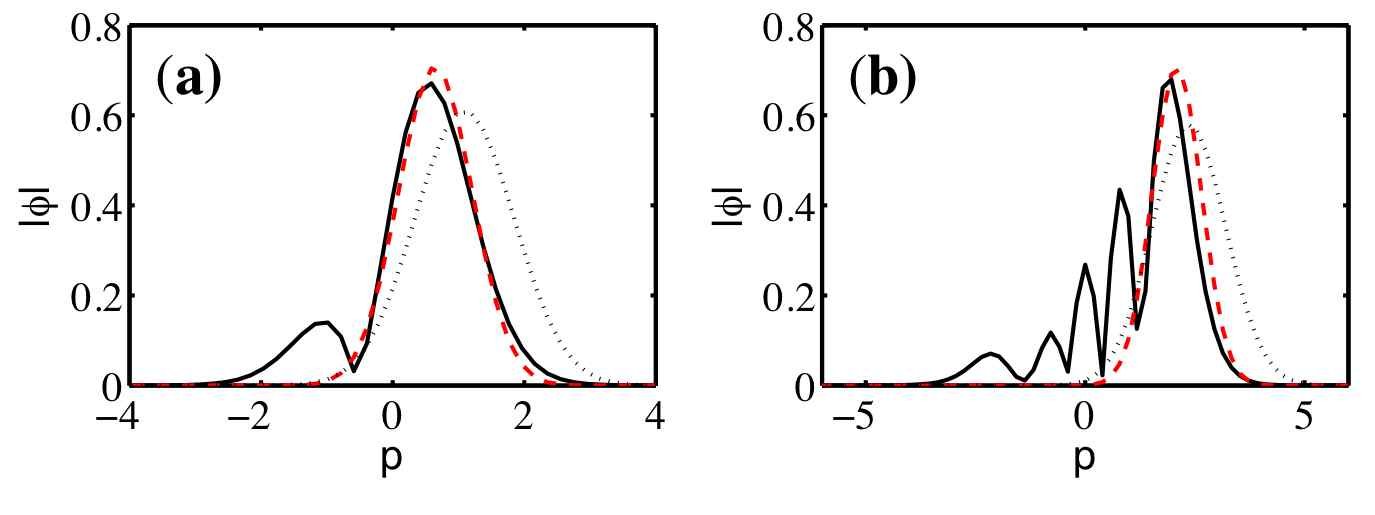},width=0.99\columnwidth} 
\caption{Experimentally realizable approximations of dynamical tunneling states. (a) Black solid line: Fourier spectrum $|\tilde{\phi}(k)|$ of exact nonlinear Floquet state $\varphi_{+}$ for $\kappa=1.3$, $\epsilon=0.2$, $\kbar=0.5$ and $U=0.012$. Black dotted line: Fourier spectrum of trap ground-state for $\sub{\kappa}{ini}=\kappa$ with initial momentum kick $p_{0}'=1.09$, where $p_{0}$ corresponds to the classical fixed point. Red dashed line: Fourier spectrum of trap ground-state for adjusted trap strength $\sub{\kappa}{ini}=0.4$ to tune the momentum space width, with initial momentum kick $p_{0}=0.65$. (b) The same as (a), but for parameters $\kappa=2$, $\epsilon=0.1$, $\kbar=0.5$ and $U=0.11$, $\sub{\kappa}{ini}=0.4$, $p_{0}'=2.37$ and $p_{0}=2.09$.
\label{initial_state_matching}}
\end{figure}

Let us first briefly review our method for matching the initial condensate closely onto the nonlinear Floquet state for a particular value of $U$ \cite{martin:matthew:chip}, using for illustration purposes the parameters $\kappa=1.3$,  $\epsilon=0.2$, $\kbar=0.5$. For these, interesting intermittent trapping behaviour has been highlighted in \rref{wuester:trappingDT} and in \fref{dyntunnel_momspace2}. In \fref{initial_state_matching}(a), we show the $t=0$ Fourier space representation of the target nonlinear Floquet state centered on the upper island of stability as a solid black line.

If we were to simply apply a momentum kick $p_{0}$ to the condensate in the ground state of the potential at $t=0$ the resulting Fourier spectrum would be given by the black dotted line. Such a kick could either be imparted by a shift of the trapping potential \cite{martin:matthew:chip} or by phase imprinting and would result in the state $\psi_{0}(x)\exp{(i p_{0}x)}$, where $\psi_{0}(x)$ is the trap ground state  and  $p_{0}$ corresponds to the classical phase-space resonance. While the resulting overlap with the target Floquet state is quite acceptable in \fref{initial_state_matching}(a), the procedure gives worse results when the Floquet state is multi-peaked and/or narrow, as in the second example, \fref{initial_state_matching}(b). In that case the simple kick reproduces the most prominent peak less well regarding its position and width. 

The situation can be improved by correctly engineering also the momentum space \emph{width} of the condensate initial state. The width 
can be tuned by preparing the atom cloud in a potential with a strength that differs from that used during the time evolution:
\begin{align}
\sub{V}{ini}(x,t<0)&=\sub{\kappa}{ini} (1 + \epsilon)\sqrt{1+ x^{2}},
\label{Vini}
\end{align}
with $\sub{\kappa}{ini}\neq\kappa$. Consequently one obtains two parameters, $\sub{\kappa}{ini}$ and $p_{0}$, with respect to which the overlap between the initial state and the target nonlinear tunneling state can be maximized. The optimal results obtained are shown for both examples in \fref{initial_state_matching} using red-dashed lines.

\begin{figure}[ht]
\centering
\epsfig{file={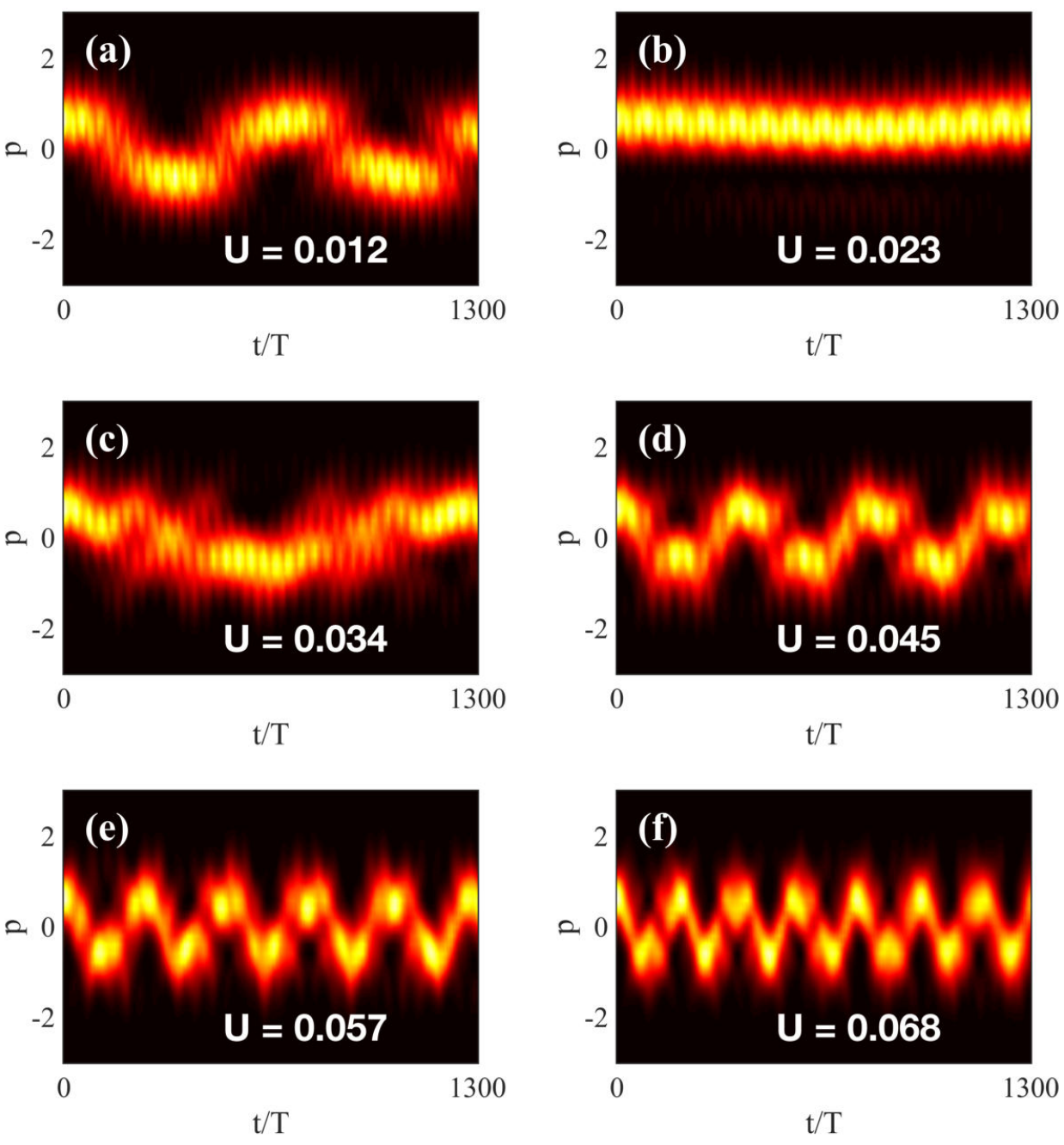},width=\columnwidth} 
\caption{Stroboscopically sampled momentum space from imperfect initial state as shown in \frefp{initial_state_matching}{a}, which approximates $\phi_{+}(x,t=0)$. $p=\kbar k$ with $\kbar=0.5$ and $\kappa=1.3$, $\epsilon=0.2$. We see intermittent trapping of tunneling oscillations that is subsequently lifted, in accordance with simulations beginning from the full nonlinear Floquet states for the same parameters, compare \fref{dyntunnel_momspace2} and \fref{twomode_coefficients2}. However the actual tunneling period for these cases deviates from our previous simulations, due to the imperfectly realised initial state.
\label{imperfect_momspace}}
\end{figure}

The case shown in \fref{initial_state_matching}(a) corresponds to $97\%$ overlap between the experimental initial state and the nonlinear Floquet state. Even with this rather crude method dynamical tunneling can now be observed in the presence of nonlinearities, generalizing earlier results from the linear case \cite{hensinger:pra}. The ensuing stroboscopic momentum space densities are displayed in \fref{imperfect_momspace}, where dynamical tunneling oscillations are clearly visible. 
However, the effective nonlinear tunneling periods are slightly different compared to \fref{dyntunnel_momspace2}. We find that the deviations depend on the precise values of $k_0$ and $\sub{\kappa}{ini}$ used to define our initial state. This will be investigated further in the next section.
\begin{figure}[ht]
\centering
\epsfig{file={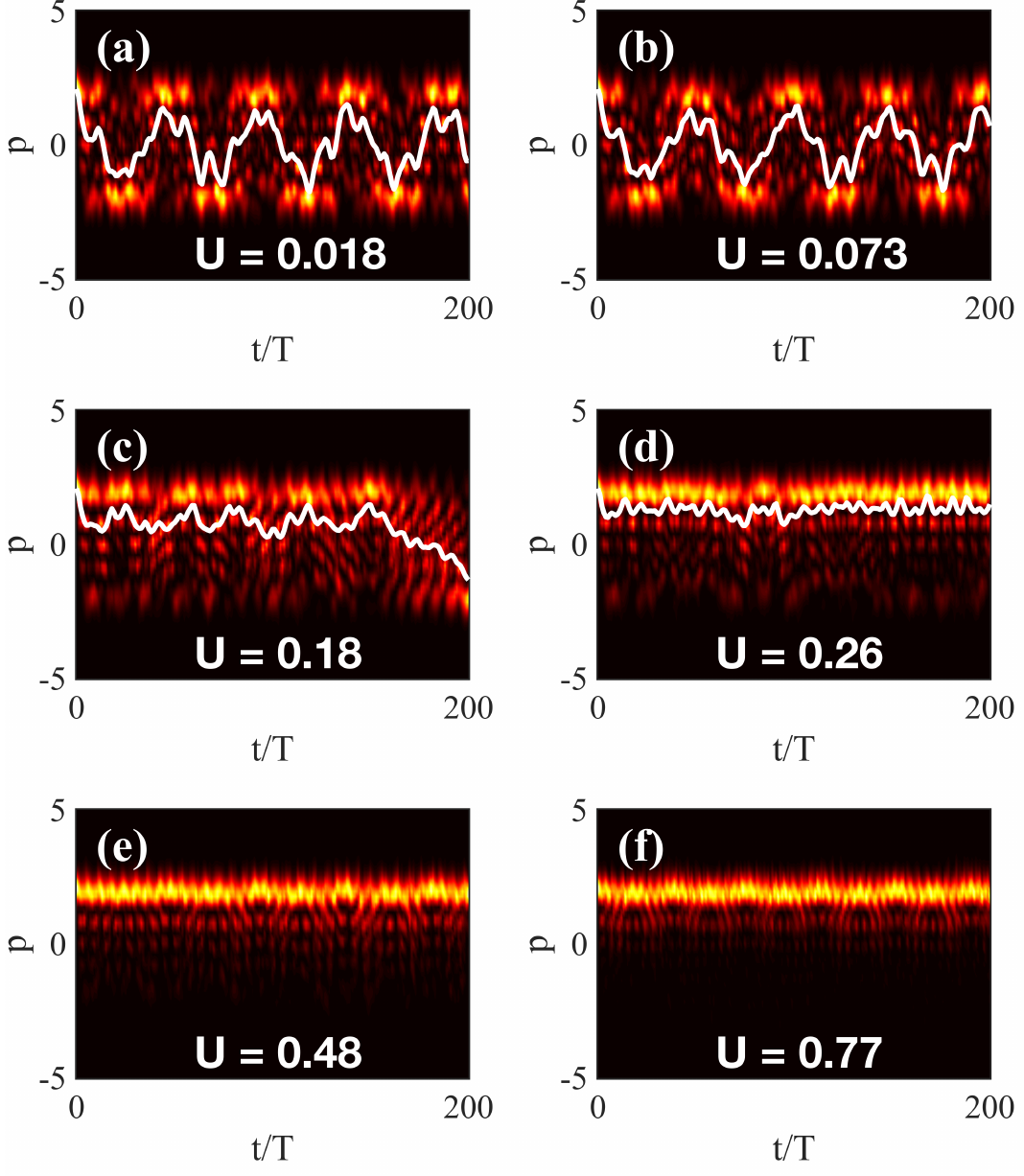},width=\columnwidth} 
\caption{Stroboscopically sampled momentum space from imperfect initial state, which approximates $\phi_{+}(x,t=0)$. $p=\kbar k$ with $\kbar=0.5$ and $\kappa=2$, $\epsilon=0.1$. The thin white line in panels (a-d) is the mean momentum. As in \fref{imperfect_momspace} dynamical tunneling persists for the case here. There is, however, a deviation in critical nonlinearity from the predicted $\sub{U}{crit}=0.079$.
\label{imperfect_momspace2}}
\end{figure}
Dynamical tunneling with an approximation of the Floquet initial state is also possible for the case of \fref{initial_state_matching}(b), as shown in \fref{imperfect_momspace2}. The initial overlap with the Floquet state is smaller, and hence deviations from two-mode behaviour are larger.

 %%%%%%%%%%%%%%%%%%%%%%%%%%%%%%%%
%%%%%%%%%%%%%%%%%%%%%%%%%%%%%%%%
\subsection{Response of tunneling period to initial-state imperfections}
\label{periodchange}

As mentioned in the previous section and in \rref{martin:matthew:chip}, the dynamical tunneling period slightly deviates from the expectations based on precise Floquet theory if the Gaussian approximations of Floquet states of \sref{floquetapprox} are used.

\begin{figure}[ht]
\centering
\epsfig{file={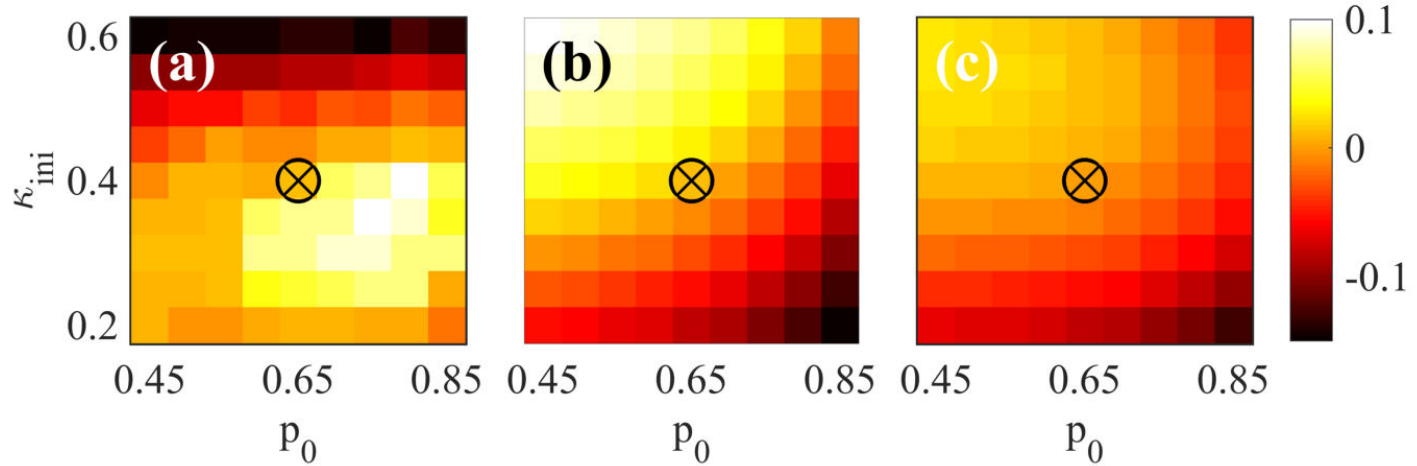},width=\columnwidth} 
\caption{ (a) Fractional change of tunneling period for varied parameters of an initial state that only approximates a nonlinear Floquet state. We construct approximate Floquet states as in \fref{initial_state_matching}(a) for the case of \fref{imperfect_momspace}(a), then vary $p_0$ and $\sub{\kappa}{ini}$ as indicated. The relative difference to the reference point, marked $\varotimes$ and with parameters as in \fref{initial_state_matching}(a), is shown as color shading. The mean period for these parameters is about $650$ driving periods. (b) Corresponding fractional change of mean nonlinear interaction energy $\sub{\bar{\qenl}}{int}$, as defined in the text. (c) Corresponding deviation $\Delta\bar{N}$ from two-mode behaviour, as defined in the text. 
\label{periodvariations}}
\end{figure}
Variations of the true dynamical tunneling period as a function of the parameters $p_0$ and $\sub{\kappa}{ini}$ used to define the initial state are shown in \fref{periodvariations}. Overall the period is rather robust against even substantial imperfections of the initial state, but its variations would have to be taken into account for the interpretation of experiments.

These changes of period can clearly be traced back to the atomic interactions as they vanish for $U=0$. 
We explored whether the modifications to the period are related to the changes in nonlinear interaction energy, $\sub{\bar{\qenl}}{int}=U\int_0^T dt \int dx |\psi(x,t)|^4/T$ using $T=15\times2\pi$ or the deviation from two-mode behaviour characterised by $\Delta\bar{N}=\int_0^T dt (|d_{+}(t)|^2+|d_{-}(t)|^2)/T - \sub{\bar{N}}{ref}$, where $\sub{\bar{N}}{ref}$ is the two-mode population for a reference data-set, indicated by the symbol $\varotimes$ in \fref{periodvariations}.
However, \fref{periodvariations}(b--c) shows that neither of these are obviously linked to the observed behaviour of the tunneling period.
The precise mechanism by which the period changes thus warrants further study. 

%%%%%%%%%%%%%%%%%%%%%%%%%%%%%%%%
%%%%%%%%%%%%%%%%%%%%%%%%%%%%%%%%
%%%%%%%%%%%%%%%%%%%%%%%%%%%%%%%%
\section{Conclusions}
\label{conclusions}

We have examined the dynamical tunneling of a Bose-Einstein condensate in a periodically modulated anharmonic potential.
We have demonstrated smooth dynamical tunneling oscillations over up to thousands of modulation periods, in the presence of nonlinear interatomic interactions. These results employ nonlinear Floquet states of the GPE.

However the interactions can have a strong influence on the tunneling. Above a critical interaction strength we have numerically found the arrest of tunneling due to macroscopic quantum self-trapping. We have derived a two-mode model with can be used to predict the critical interaction strength $\sub{U}{crit}$ from knowledge of the linear Floquet states. It is also possible to accurately estimate the evolution of the atomic population in the tunneling states involved. The performance of this model and empirical values for $\sub{U}{crit}$ were surveyed as a function of parameters $\kappa$,$\epsilon$ and $\kbar$.

Finally we have investigated nonlinear dynamical tunneling commencing from an experimentally realistic initial state rather than  the actual Floquet state. Such initial states can be prepared by imparting a momentum-kick and adjusting the trap strength, and thus are experimentally more tractable than the mathematical Floquet state. We find that the imperfect state yields dynamical tunneling and macroscopic self-trapping, with only minor deviations in observed dynamical tunneling periods.

%%%%%%%%%%%%%%%%%%%%%%%%%%%%%%%%
\acknowledgments
We would like to thank P.~B.~Blakie for help with conjugate gradient routines, M.~Lenz for code to obtain and diagonalize the linear Floquet operator and S. Holt and B.~J.~D\c{a}browska-W{\"u}ster for their contribution to the initial phase of this project. This research was supported by the Australian Research Council Centre of Excellence in Future Low-Energy Electronics and Technologies (FLEET, Project No.~CE170100039), and the Australian government Department of Industry, Science,
and Resources via the Australia-India Strategic Research
Fund (AIRXIV000025).
It was also supported by the Department of Science and Technology (DST), Government of India under Australia India Strategic
Research Funds (DST/INT/AUS/P-84/2022).

%%%%%%%%%%%%%%%%%%%%%%%%%%%%%%%%
\appendix
%%%%%%%%%%%%%%%%%%%%%%%%%%%%%%%%

%%%%%%%%%%%%%%%%%%%%%%%%%%%%%%%%
%%%%%%%%%%%%%%%%%%%%%%%%%%%%%%%%
\section{Atom chip potential}
\label{atomchip}

In Ref.~\cite{martin:matthew:chip} we discuss how the radial direction of a typical atom-chip potential
allows one to engineer a one-dimensional potential of the form:
\begin{align}
V(x,t)&=V_{0} \big[ 1 + \epsilon \cos(\Omega t) \big] \big[ \sqrt{1+ (x/d)^{2}  } - 1 \big],
\label{chip_potential_appendix}
\end{align}
where $x$ is the spatial coordinate, $V_{0}$ is the potential strength, $d$ describes its width and $\epsilon$ controls the strength of the modulation with frequency $\Omega$. 

We assume that the Bose-Einstein condensate is well described by the one-dimensional Gross-Pitaevskii equation:
\begin{align}
i\hbar \frac{\partial}{\partial t} \psi& =\left[ -\frac{\hbar^{2}}{2m}\frac{\partial^{2}}{\partial x^{2}} + V(x,t) + \sub{\gamma}{1D} |\psi|^{2} \right]\psi.
\label{1dgpe}
\end{align}
The atoms have mass $m$ and s-wave scattering length $a_{s}$, then $\sub{\gamma}{1D}=2 a_{s}/a_{\perp}$  where $a_{\perp}=\sqrt{\hbar/m\omega_{\perp}}$ is the oscillator length of tight transverse confinement with strength $\omega_{\perp}$. We assume the Gross-Pitaevskii wave function $\psi$ in \eref{1dgpe} to be normalized to the number of atoms $\int |\psi|^{2} dx=N$.

We now measure length in units of $d$, time in units of $\Omega^{-1}$ and employ the dimensionless wavefunction $\tilde{\psi}= \psi \sqrt{d/N}$, which is normalized to one. Finally relabelling $\tilde{\psi}\rightarrow \psi$ we obtain:
\begin{align}
i\kbar \frac{\partial}{\partial t} \psi& =\left[ -\frac{\kbar^{2}}{2}\frac{\partial^{2}}{\partial x^{2}} + V(x,t) + U |\psi|^{2} \right]\psi,
\label{standardgpe_appendix}
\end{align}
with
\begin{align}
V(x,t)&=\kappa \big[1 + \epsilon \cos(t) \big]\big[\sqrt{1+ x^{2}} - 1\big].
\label{dimlesspotential_appendix}
\end{align}
Here $\kappa=V_{0}/m\Omega^{2}d^{2}$, $U=N \gamma /m \Omega^{2}d^{3}$ and $\kbar=\hbar/d^{2}m \Omega$. The commutator of the position operator $\hat{x}$ and the momentum operator $\hat{p}=i\kbar \partial/\partial x$ fulfills $[\hat{x},\hat{p}]=i \kbar$, thus $\kbar$ controls the relative importance of quantum effects for the system. 

%%%%%%%%%%%%%%%%%%%%%%%%%%%%%%%%
\section{Determining nonlinear Floquet states}
\label{nonlinear_numerics}
To find the nonlinear Floquet states solving \eref{nonlin_Floqeqn}, we make use of existing methods for the solution of the \emph{time-independent} GPE in two dimensions. To this end we effectively treat time as a second spatial dimension $y$, and aim to find solutions of the following operator eigenvalue equation in two dimensions:
\begin{align}
&\left[-\frac{\kbar^{2}}{2m}\frac{\partial^{2}}{\partial x^{2}} + V(x,y) + U | \phi_{0,1}(x,y) |^{2} -i\kbar \frac{\partial}{\partial y} \right]
\phi_{0,1}(x,y)
\CR
&=\qenl_{0,1}(U) \phi_{0,1}(x,y).
\label{effective2d}
\end{align}
The solutions $\phi_{0,1}(x,y)$ are required to obey periodic boundary conditions in the time $(y)$ direction: $\phi(x,y)=\phi(x,y+2\pi)$ and to vanish for $x\rightarrow \pm X_{max}$, where $X_{max}$ is the extension of our numerical grid.

To solve \eref{effective2d} we employ conjugate gradient techniques \cite{blair:thesis}, which are efficient at determining nonlinear eigenstates if they are supplied with a good initial guess. We find nonlinear Floquet tunneling states with the following numerical scheme: 
\begin{enumerate}[(i)]
\item We construct the time evolution operator for $U=0$ and find the \emph{linear} Floquet states $|u_{n}\rangle$.
\item Among these, we identify the odd and even tunneling states $|u_{0,1}\rangle$ as those having the largest overlap with a coherent state centered on the period-one fixed point.
\item We evolve these over one period with the Schr\"odinger equation to obtain $|\chi_{0,1}(x,t)\rangle$, from which we extract $|u_{0,1}(x,t)\rangle$ using \eref{floquet_theorem}. 
\item We obtain nonlinear solutions of \eref{effective2d} for small $U$ using a conjugate gradient technique with the initial guess taken as linear result for $u_{0,1}(x,t)$ with $t\rightarrow y$.
\item We obtain tunneling states for $U+\Delta U$, using the result for $U$ as initial guess. This step is iterated until the desired $\sub{U}{max}$ is reached.
\item The states obtained are validated by evolution with the GPE (\ref{standardgpe}) over many modulation periods. 
\end{enumerate}
The nonlinear Floquet states employed in our work were obtained on $128\times200$ grid points in the $x$ and $t$ directions respectively, and verified on larger grids. Typical steps for the iteration of the nonlinearity were $\Delta U=0.05$. 

%%%%%%%%%%%%%%%%%%%%%%%%%%%%%%%%
\section{Two-mode model}
\label{twomm_appendix}

As outlined in \sref{twomm} our two-mode model is based on the odd and even nonlinear Floquet states, denoted $\phi_{0}$ (even) and $\phi_{1}$ (odd). The modes localised on the upper and lower islands are $\phi_{+}=\frac{1}{\sqrt{2}}(\phi_{0}+i \phi_{1})$ (upper) and $\phi_{-}=\frac{1}{\sqrt{2}}(\phi_{0}-i \phi_{1})$ (lower). Due to the orthogonality of $\phi_{0}$ and $\phi_{1}$ these modes are also orthogonal. We know that the modes $\phi_{n}$ $n\in{0,1}$ solve the nonlinear Floquet equation:
\begin{align}
\left(\hat{H}_{0}(x,t) + U | \phi_n(x,t)|^{2} -i\kbar \frac{\partial }{\partial t}\right)\phi_n(x,t)&=\qenl_{n}\phi_n(x,t).
\label{NLFloquet_eqn_appendix}
\end{align}
Let us now assume that the dynamics are restricted to the upper and lower island modes  $\phi_{\pm}(x,t)$, i.e.:
\begin{align}
\psi(x,t)&=c_{+}(t)\phi_{+}(x,t) + c_{-}(t)\phi_{-}(x,t) .
\label{gpeexpansion_appendix}
\end{align}
We can alternatively expand in the odd and even tunneling Floquet states:
\begin{align}
\psi(x,t)&=c_{0}(t)\phi_{0}(x,t) + c_{1}(t)\phi_{1}(x,t) .
\label{gpeexpansion2_appendix}
\end{align}
The complex coefficients are related by:
\begin{align}
c_{+}&=\frac{1}{\sqrt{2}}(c_{0}-ic_{1}),\:\:\:\:\:\:c_{-}=\frac{1}{\sqrt{2}}(c_{0}+ic_{1}),
\CR
c_{0}&=\frac{1}{\sqrt{2}}(c_{+}+c_{-}),\:\:\:\:\:\:c_{1}=\frac{i}{\sqrt{2}}(c_{+}-c_{-}),
\label{basischange}
\end{align}
The restricted expansion (\ref{gpeexpansion_appendix}) is well justified for the cases presented here. This allows us to assume:
\begin{align}
|c_{0}(t)|^{2}+|c_{1}(t)|^{2}=|c_{+}(t)|^{2}+|c_{-}(t)|^{2}&\approx1.
\label{normalis}
\end{align}
Let us write the GPE in the form
\begin{align}
\left(\hat{H}_{0}(x,t) + U |\psi(x,t)|^{2} -i\kbar \frac{\partial }{\partial t}\right)\psi(x,t)&=0.
\label{GPE}
\end{align}
Inserting \eref{gpeexpansion2_appendix} we obtain
\begin{align}
\sum_{n=0,1} &\bigg\{\bigg[\hat{H}_{0} 
+ U\big(|c_{0}|^{2}|\phi_{0}|^{2}  +|c_{1}|^{2}|\phi_{1}|^{2} 
\CR
&+ c_{0}^{*}c_{1}\phi_{0}^{*} \phi_{1}+ c_{0}c_{1}^{*} \phi_{0} \phi_{1}^{*} \big)\bigg]c_{n}\phi_{n}
\CR
 &-i \kbar \left(\frac{\partial  c_{n}}{\partial t} \right)\phi_{n}  -i\kbar c_{n}\frac{\partial }{\partial t} \phi_{n}\bigg\}=0.
\label{gpe_inserted}
\end{align}
Using \eref{normalis} and \eref{NLFloquet_eqn_appendix} we can then write:
\begin{align}
\sum_{n=0,1} &\bigg[\qenl_{n} 
+ U |c_{|n-1|}|^{2}\left(|\phi_{|n-1|}|^{2} - |\phi_{n}|^{2} \right)
\label{gpe_inserted2}
\nnl
&+ U\left(c_{0}^{*}c_{1}\phi_{0}^{*} \phi_{1}+ c_{0}c_{1}^{*} \phi_{0} \phi_{1}^{*} \right)
-i \frac{\kbar}{c_n} \left(\frac{\partial  c_{n}}{\partial t} \right) \bigg] c_{n}\phi_{n}=0.
\end{align}
For this step we have inserted $0$ in the form of $|c_{1}|^{2}|\phi_{0}|^{2}-|c_{1}|^{2}|\phi_{0}|^{2}$ into the $n=0$ term of the sum and the corresponding expression with $1\leftrightarrow0$ into the $n=1$ term. Now we project this result onto $\phi_{m}$, $m\in0,1$ and obtain:

\begin{align}
&i\kbar \frac{\partial }{\partial t} c_{m}=
\qenl_{m} c_{m} + U \sum_{n=0,1}\bigg[ 
\CR
& c_{0}^{*}c_{1} \langle \phi_{m}|\phi_{0}^{*}\phi_{1}|\phi_{n}\rangle c_{n} 
+ c_{0}c_{1}^{*} \langle \phi_{m}|\phi_{0}\phi_{1}^{*}|\phi_{n}\rangle c_{n} 
\CR
&+ |c_{|n-1|}|^{2}\
\langle \phi_{m}|\left(|\phi_{|n-1|}|^{2} - |\phi_{n}|^{2} \right)|\phi_{n}\rangle c_{n} 
\bigg].
\label{gpe_projected}
\end{align}
Of the overlap integrals appearing in \eref{gpe_projected}, many vanish due to odd symmetry of the integrand. We label the remaining overlap integrals:
\begin{subequations}
\label{couplings_appendix}
\begin{align}
U_{01}(t)&=U\int dx\: |\phi_{0}(x,t)|^{2}|\phi_{1}(x,t)|^{2},
\\
A_{01}(t)&=U\int dx\: \phi_{0}(x,t)^{2}\phi_{1}(x,t)^{*2},
\\
U_{00}(t)&=U\int dx\: |\phi_{0}(x,t)|^{4},
\\
U_{11}(t)&=U\int dx\: |\phi_{1}(x,t)|^{4}.
\end{align}
\end{subequations}
With this notation we can finally write the coupled equations for the coefficents of the odd and even states as:
\begin{subequations}
\begin{align}
i\kbar \frac{\partial }{\partial t} c_{0}&=\qenl_{0}c_{0} +A_{01}^{*} c_{1}^{2}c_{0}^{*} +U_{01}|c_{1}|^{2} c_{0} 
\CR
&+ (U_{01} - U_{00}) |c_{1}|^{2} c_{0},
\\
i\kbar \frac{\partial }{\partial t} c_{1}&=\qenl_{1}c_{1} +A_{01} c_{0}^{2}c_{1}^{*} +U_{01}|c_{0}|^{2} c_{1}
\CR
&+ (U_{01} - U_{11}) |c_{0}|^{2} c_{1}.
\label{oddeveneqns}
\end{align}
\end{subequations}
At this stage we convert from the odd-even coefficients to the up-down coefficients, using \eref{basischange}. 
\begin{subequations}
\label{updowneqnsboth}
\begin{align}
\label{updowneqns}
i\kbar \frac{\partial }{\partial t} c_{+}
&=\bar{\qenl}c_{+} + \frac{\Delta \qenl}{2}c_{-} + {\mathcal N}[c_{+},c_{+}^{*},c_{-},c_{-}^{*},t]  
\end{align}
\begin{align}
\label{updowneqns2}
i\kbar \frac{\partial }{\partial t} c_{-}
&=\bar{\qenl}c_{-} + \frac{\Delta \qenl}{2}c_{+} +  {\mathcal N}[c_{-},c_{-}^{*},c_{+},c_{+}^{*},t]
\end{align}
\end{subequations}
where $\bar{\qenl}=(\qenl_{0} + \qenl_{1})/2$ and $\Delta \qenl=\qenl_{0} - \qenl_{1}$ and:
\begin{align}
	&{\mathcal N}[a,b,c,d,t] = \left[U_{01} -\frac{\real{A_{01}}}{2} \right]a^2b
\CR
&-\left[U_{01} + \frac{\real{A_{01}}}{2} \right]c^2b
+\real{A_{01}}acd
\CR
&+i \imag{A_{01}}\left[\frac{c^2d}{2} +\frac{a^2d}{2} - abc\right]
\CR
&+ \frac{U_{00}}{4}\left[
a^2d + c^2 b - c^2d - a^2b
\right]
\CR
&-\frac{U_{11}}{4}\left[
a^2b - c^2d + a^2d - c^2b
\right].
	\label{nonlinear_terms}
\end{align}
As explained in \sref{twomm}, one can proceed by either extracting the full time-dependent coupling coefficients such as $U_{01}(t)$ from the nonlinear Floquet states, or alternatively using coefficients averaged over one period: $U_{ij}(t)\rightarrow \bar{U}_{ij}=\int_{0}^{T}U_{ij}(t)dt/T$.
Here we provide some additional detail on both methods.

\begin{figure}[ht]
\centering
\epsfig{file={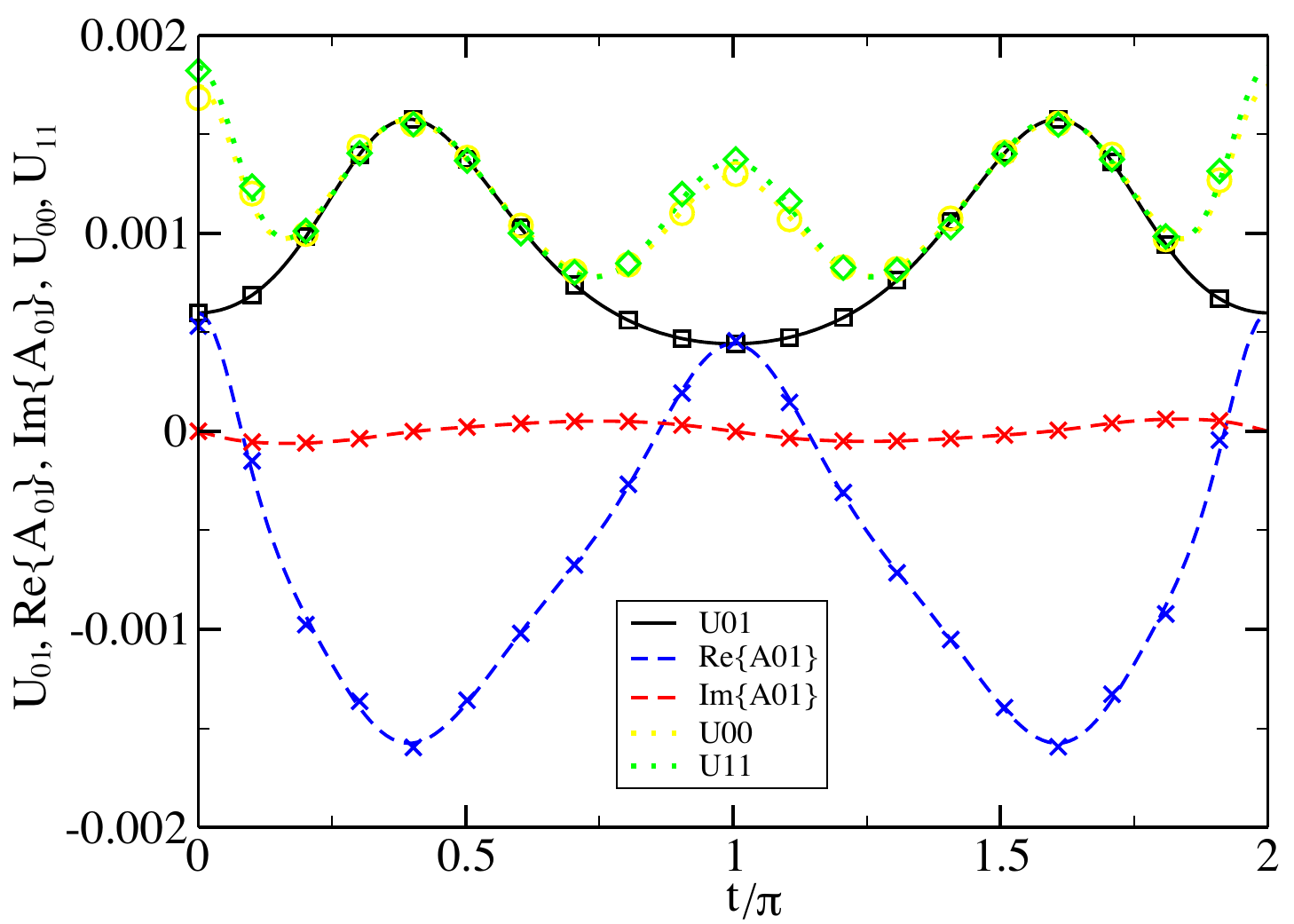},width=\columnwidth} 
\caption{Time dependence of two-mode theory coupling coefficients over one period. (black, solid) $U_{01}$, (blue, dashed) $\real{A_{01}}$, (red, dashed) $\imag{A_{01}}$ , (green, dottted) $U_{00}$, (yellow, dotted) $U_{11}$. The markers of the corresponding colour show the quality of approximating these coefficients by 9 harmonics. We find that using average values of the coefficients usually gives as good tunneling results as the full form. Note that $\imag{A_{01}}$ always averages to zero.
\label{tdepcouplings}}
\end{figure}

For the explicit method (without averaging) we proceed as follows. We identify the 9 most prominent harmonics of the periodic functions in \eref{couplings_appendix} and use these to obtain an analytical approximation that can be used for an adaptive time-step numerical solution of Eqs.~\ref{updowneqns}-\ref{updowneqns2}. The evolution of the coupling coefficients over one period is shown for an example in \fref{tdepcouplings}. 

Using averaged values for the coefficients in \fref{tdepcouplings} is justified when the population oscillations of $c_{+}$ and $c_{-}$ take place on time scales much larger than a single period. In almost all cases that we studied we found the more accurate but much more involved explicit method unnecessary and using averaged coefficients gave good results. One further analytical simplification is possible in this case: It can be shown that $\imag{\bar{A}_{01}}=0$.  
From \eref{couplings_appendix} we can expand 
\begin{align}
\label{Abar01zeroproof}
\imag{\bar{A}_{01}}&=\frac{2}{T}\int_0^Tdt \int_{-\infty}^{\infty}dx\big[\phi_{0}(x,t)''^2\phi_{1}(x,t)'\phi_{1}(x,t)''
\CR
&+\phi_{0}(x,t)'\phi_{0}(x,t)''\phi_{1}(x,t)'^2 
\CR
&- \phi_{0}(x,t)'\phi_{0}(x,t)''\phi_{1}(x,t)''^2
\CR
&-\phi_{0}(x,t)'^2\phi_{1}(x,t)'\phi_{1}(x,t)''\big],
\end{align}
where a prime denotes the real part, and a double prime the imaginary part.
The time average can be replaced by $\int_0^T dt/T \rightarrow \int_{-T/2}^{T/2} dt/T$ due to the periodicity of the Floquet states. Inspection of \eref{effective2d} shows that $\phi_{0,1}(t)=\phi_{0,1}^*(-t)$, hence $\phi_{0,1}(t)'=\phi_{0,1}(-t)'$ and $\phi_{0,1}(t)''=-\phi_{0,1}(-t)''$. From these symmetry properties all four terms in \eref{Abar01zeroproof} vanish when averaged over time.

As stated in \sref{twomm}, the main advantage of averaging the coefficients is that we can further proceed with the analysis of the equations following \rref{smerzi:mqst}. We transform to a description in terms of the population imbalance $z=N_{+}-N_{-}$, and relative phase $\varphi=\theta_{-}-\theta_{+}$ where  $c_{\pm}=\sqrt{N_{\pm}}e^{i\theta_{\pm}}$ with $N_{\pm}, \theta_{\pm} \in \mathbb{R}$. Note that with this normalization $N_{+}+N_{-}=1$. 

The dynamical equations for $z$ and $\varphi$ that can be derived from Eqs.~\ref{updowneqns}-\ref{updowneqns2}, using  $\imag{\bar{A}_{01}}=0$, are:
\begin{align}
\label{zeqn}
\kbar \frac{\partial }{\partial t} z&=\left[\Delta \qenl +\frac{\bar{U}_{11}-\bar{U}_{00} }{2} \right]
\sqrt{1-z^{2}}\sin{(\varphi)}
\nnl
&+\left[\frac{ \bar{U}_{00}+\bar{U}_{11} }{4} -\bar{U}_{01} -\frac{ \real{\bar{A}_{01}} }{2} \right](1-z^{2})\sin{(2\varphi)},
\end{align}
\begin{align}
\label{phieqn}
\kbar \frac{\partial }{\partial t} \varphi&=-\left[\Delta \qenl +\frac{\bar{U}_{11}-\bar{U}_{00} }{2} \right]
\frac{z}{\sqrt{1-z^{2}} }\cos{(\varphi)}
\nnl
&-\left[\frac{ \bar{U}_{00}+\bar{U}_{11} }{4} -\bar{U}_{01} -\frac{ \real{\bar{A}_{01}} }{2} \right]z\cos{(2\varphi)},
\nnl
&-\left[\frac{ \bar{U}_{00}+\bar{U}_{11} }{4} -\bar{U}_{01} +\frac{3}{2} \real{\bar{A}_{01}} \right]z.
\end{align}
These equations are Euler-Lagrange equations for canonical variables $z$, $\varphi$ of the Hamiltonian:
\begin{align}
\sub{H}{eff}&=\frac{1}{2} \Lambda z^{2} + \alpha \sqrt{1-z^{2} }\cos{(\varphi)} + \beta (1-z^{2})\cos{(2\varphi)},\label{effhamiltonian_appendix}
\end{align}
with parameters:
\begin{subequations}
\begin{align}
\qe&=\frac{U_{00}+U_{11} }{4} + \frac{3}{2} \real{A_{01}}-U_{01},
\\
\alpha&=\frac{U_{00}-U_{11} }{2} - \Delta \qenl,
\\
\beta&=\frac{U_{01}}{2} -\frac{U_{00}+U_{11} }{8}+ \frac{ \real{A_{01} } }{4}.
\end{align}
\end{subequations}
As outlined in \sref{twomm}, the effective Hamiltonian allows us to derive a condition for the shutdown of tunneling oscillations:
\begin{align}
\frac{1}{2}\Lambda&=\alpha \cos{(\varphi(t))} +\beta \cos{(2\varphi(t))},
\label{maxcoupling1_appendix}
\end{align}
An approximative treatment of this equation was given in \sref{twomm}. For some of the cases surveyed in \sref{parametersurvey} it is not sufficient, and can instead be analyzed with the following method. Let $f(\varphi)=\alpha \cos{(\varphi)} +\beta \cos{(2\varphi)}$. The values of $f$ at its extrema are $\alpha+\beta$, $-\alpha+\beta$ and $-\alpha^2/8\beta -\beta$. Self trapping occurs whenever $\Lambda$ exceeds the global maximum of $f$, or lies below the minimum. Precisely which is the condition for self trapping depends on the signs of $\Lambda$, $\alpha$, $\beta$ and of $|4\beta| -\alpha$. These in turn depend on $U$. To predict $\sub{U}{crit}$ from the knowledge of the linear Floquet states more accurately than by using \eref{critical_coupling}, we found it most convenient to numerically determine the largest $U$ for which
\begin{align}
\frac{1}{2}U \Lambda_{0}&=\left(-\Delta \qenl +U \frac{\bar{O}_{00} - \bar{O}_{11}}{2}\right) \cos{(\varphi(t))}
\CR 
&+ U \beta_{0} \cos{(2\varphi(t))}
\label{maxcoupling1_appendix2}
\end{align}
still has a solution for $\varphi$. We denote this value with $\sub{U}{c,full}$. Here the parameters $\Lambda_{0}=\Lambda/U$, $\beta_{0}=\beta/U$ and $\bar{O}_{ij}=\bar{U}_{ij}/U$ are independent of $U$.

%%%%%%%%%%%%%%%%%%%%%%%
\bibliography{dyntunnel}

\begin{thebibliography}{69}
\bibitem [{foo()}]{footnote:isolated_values}%
  \BibitemOpen
  \href@noop {} {}\bibinfo {note} {For some parameters the set of discrete
  nonlinearities $\{U_j\}$ contained single values of $U_k$ that give rise to
  suppressed tunnelling. These were not identified as ``intermittently trapped"
  for this figure.}\BibitemShut {Stop}%
\end{thebibliography}
\end{document}